\documentclass{jedm}

\usepackage{booktabs} 
\usepackage{ftnxtra}
\usepackage{fnpos}
\usepackage{lmodern}
\usepackage{amsmath}
\usepackage{amsfonts}
\usepackage{array}
\usepackage[caption=false,font=footnotesize]{subfig}
\usepackage{stfloats}
\usepackage{url}
\usepackage{multirow}
\usepackage[normalem]{ulem}
\usepackage{threeparttable}
\usepackage{hyperref}
\usepackage{booktabs}
\usepackage{algpseudocode}
\usepackage{algorithm}
\usepackage{amsthm}
\usepackage{bm}
\usepackage{graphicx}
\usepackage{color}
\usepackage{subfig}
\DeclareGraphicsExtensions{.png,.pdf}

\newcommand{\ul}{\underline}
\newcommand{\model}{Course2vec}
\newcommand{\modelplusplus}{Course2vec(++)}
\newcommand{\modelplus}{Course2vec(+)}
\newcommand{\modelplusminus}{Course2vec(+-)}

\newcommand{\grppopplusminus}{grp-pop(+-)}

\newcommand{\goodrecall}{Recall(good)}
\newcommand{\badrecall}{Recall(bad)}
\newcommand{\diffrecall}{Recall(diff)}

\newcommand{\seq}{\mbox{$Q$}}
\newcommand{\term}{\mbox{$\mathcal{T}$}}
\newcommand{\R}{\mbox{$\mathbb{R}$}}
\newcommand{\inmat}{\mbox{$\mathbf{W}$}}
\newcommand{\outmat}{\mbox{$\mathbf{W'}$}}

\newcommand{\outvec}{\mbox{$\mathbf{w'}$}}
\newcommand{\hidden}{\mbox{$\mathbf{h}$}}
\newcommand{\students}{\mbox{$\mathcal{S}$}}
\newcommand{\prev}{\mbox{$\mathcal{P}$}}
\newcommand{\good}{\mbox{$\mathcal{G}$}}
\newcommand{\bad}{\mbox{$\mathcal{B}$}}

\newcommand{\freqmatplusplus}{\mbox{$\mathbf{F^{++}}$}}
\newcommand{\freqmatplus}{\mbox{$\mathbf{F^+}$}}
\newcommand{\freqmatplusminus}{\mbox{$\mathbf{F^{+-}}$}}

\newcommand{\gradescore}{\mbox{$\hat{g}$}}
\newcommand{\recscore}{\mbox{$\hat{r}$}}
\newcommand{\finalscore}{rank-score}

\newcommand{\curgoodgpa}{cur-good}
\newcommand{\curbadgpa}{cur-bad}
\newcommand{\prevgpa}{prev-gpa}

\newcommand{\red}{\textcolor{black}}
\newcommand{\blue}{\textcolor{black}}

\begin{document}
\title{Will this Course Increase or Decrease Your GPA? Towards Grade-aware
  Course Recommendation\footnote{An early version of this paper is published as
  a technical report here: \url{https://goo.gl/HrxVdr}.}}
\date{}

\author{{\large Sara Morsy}\\
  Department of Computer Science\\
  and Engineering\\
  University of Minnesota\\
  morsy@cs.umn.edu \and
  {\large George Karypis}\\
  Department of Computer Science\\
  and Engineering\\
  University of Minnesota\\
  karypis@cs.umn.edu}

\maketitle

\begin{abstract}
  In order to help undergraduate students towards successfully completing their
  degrees, developing tools that can assist students during the course
  selection process is a significant task in the education domain. The optimal
  set of courses for each student should include courses that help him/her
  graduate in a timely fashion and for which he/she is well-prepared for so as
  to get a good grade in. To this end, we propose two different
  \emph{grade-aware course recommendation} approaches to recommend to each
  student his/her optimal set of courses. The first approach ranks the courses
  by using an objective function that differentiates between courses that are
  expected to increase or decrease a student's GPA. The second approach
  combines the grades predicted by grade prediction methods with the rankings
  produced by course recommendation methods to improve the final course
  rankings. To obtain the course rankings in the first approach, we adapt two
  widely-used representation learning techniques to learn the optimal temporal
  ordering between courses. Our experiments on a large dataset obtained from
  the University of Minnesota that includes students from 23 different majors
  show that the grade-aware course recommendation methods can do better on
  recommending more courses in which the students are expected to perform well
  and recommending fewer courses in which they are expected not to perform well
  in than grade-unaware course recommendation methods.

\end{abstract}

\section{Introduction}

The average six-year graduation rate across four-year higher education
institutions has been around 59\% over the past 15
years~\cite{kena2016condition,braxton2011understanding}, while less than half
of college graduates finish within four
years~\cite{braxton2011understanding}. These statistics pose challenges in
terms of workforce development, economic activity and national
productivity. This has resulted in a critical need for analyzing the available
data about past students in order to provide actionable insights to improve
college student graduation and retention rates. Some examples of the problems
that have been investigated are: course
recommendation~\cite{elbadrawy2016domain,bendakir2006using,lee2011intelligent,parameswaran2009recommendations,parameswaran2010recsplorer,parameswaran2010evaluating,parameswaran2011recommendation},
next-term course grade
prediction~\cite{polyzou2016grade,sweeney2016next,elbadrawy2016domain,morsy2017cumulative,hu2018course},
predicting the final grade of the course based on the student's ongoing
performance during the term~\cite{meier2015personalized}, in-class activities
grade prediction~\cite{elbadrawy2015collaborative}, predicting student's
performance in tutoring
systems~\cite{thai2011factorization,Hershkovitz2013predicting,hwang2015unified,romero2008data,thai2012using},
and knowledge tracing and student
modeling~\cite{reddy2016latent,lan2014sparse,gonzalez2012dynamic}.

Both \emph{course
  recommendation}~\cite{bendakir2006using,parameswaran2011recommendation,elbadrawy2016domain,bhumichitr2017recommender,hagemann2018module}
and \emph{grade
  prediction}~\cite{sweeney2016next,elbadrawy2016domain,polyzou2016grade,morsy2017cumulative,hu2018course}
methods aim to help students during the process of course registration in each
semester. By learning from historical registration data, course recommendation
focuses on recommending courses to students that will help them in completing
their degrees. Grade prediction focuses on estimating the students' expected
grades in future courses. Based on what courses they previously took and how
well they performed in them, the predicted grades give an estimation of how
well students are prepared for future courses. Nearly all of the previous
studies have focused on solving each problem separately, though both problems
are inter-related in the sense that they both aim to help students graduate in
a timely and successful manner.

In this paper, we propose a \emph{grade-aware course recommendation} framework
that focuses on recommending a set of courses that will help students: (i)
complete their degrees in a timely fashion, and (ii) maintain or improve their
GPA. To this end, we propose two different approaches for recommendation. The
first approach ranks the courses by using an objective function that
differentiates between courses that are expected to increase or decrease a
student's GPA. The second approach uses the grades that students are expected
to obtain in future courses to improve the ranking of the courses produced by
course recommendation methods.

To obtain course rankings in \blue{the first approach}, we adapt two
widely-known representation learning techniques, which have proven successful
in many fields, to solve the grade-aware course recommendation problem. The
first is based on Singular Value Decomposition (SVD), which is a linear model
that learns a low-rank approximation of a given matrix. The second, which we
refer to as {\model}, is based on Word2vec~\cite{mikolov2013efficient} that
uses a log-linear model to formulate the problem as a maximum likelihood
estimation problem. In both approaches, the courses taken by each student are
treated as temporally-ordered sets of courses, and each approach is trained to
learn these orderings.

\subsection{Contributions}

The main contributions of this work are the following:
\begin{enumerate}

\item We propose a \emph{Grade-aware Course Recommendation} framework in higher
  education that recommends courses to students that the students are most
  likely to register for in their following terms and that will help maintain
  or improve their overall GPA. The proposed framework combines the benefits of
  both course recommendation and grade prediction approaches to better help
  students graduate in a timely and successful manner.

\item We investigate two different approaches for solving grade-aware course
  recommendation. The first approach uses an objective function that explicitly
  differentiates between good and bad courses, while the other approach
  combines grade prediction methods with course recommendation methods in a
  non-linear way.

\item We adapt two-widely used representation learning techniques to solve the
  grade-aware course recommendation problem, by modeling historical course
  ordering data and differentiating between courses that increase or decrease
  the student's GPA.

\item We perform an extensive set of experiments on a dataset spanning 16 years
  obtained from the University of Minnesota, which includes students who belong
  to 23 different majors. The results show that: (i) the proposed grade-aware
  course recommendation approaches outperform grade-unaware course recommendation
  methods in recommending more courses that increase the students' GPA and fewer
  courses that decrease it; and (ii) the proposed representation learning
  approaches outperform competing approaches for grade-aware course
  recommendation in terms of recommending courses which students are expected to
  perform well in, as well as differentiating between courses which students are
  expected to perform well in and those which they are expected not to perform
  well in.

\item \red{We provide an in-depth analysis of the recommendation accuracy
    across different majors and different student groups. We show the
    effectiveness of our proposed approaches on different majors and student
    groups over the best competing method. In addition, we analyze two
    important characteristics for the recommendations: the course difficulty as
    well as the course popularity. We show that our proposed approaches are not
    prone to recommending easy courses. Furthermore, they are able to recommend
    courses with different popularity in a similar manner.}
\end{enumerate}



\section{Related Work}
\label{sec:related}

\subsection{Course Recommendation}
\label{sec:related:course-rec}

Different machine learning methods have been recently developed for course
recommendation. For example, ~\citeN{bendakir2006using} used association rule
mining to discover significant rules that associate academic courses from
previous students' data. ~\citeN{lee2011intelligent} ranked the courses
for each student based on the course's importance within his/her major, its
satisified prerequisites, and the extent by which the course adds to the
student's knowledge state.

Another set of recommendation methods proposed
in~\cite{parameswaran2009recommendations,parameswaran2010recsplorer,parameswaran2010evaluating,parameswaran2011recommendation}
focused on satisfying the degree plan's requirements that include various
complex constraints. The problem was shown to be NP-hard and different
heuristic approaches were proposed in order to solve the problem.

~\citeN{elbadrawy2016domain} proposed using both
student- and course-based academic features, in order to improve the
performance of three popular recommendation methods in the education domain,
namely: popularity-based ranking, user-based collaborative filtering and
matrix factorization. These features are used to define finer groups of
students and courses and were shown to improve the recommendation performance
of the three aforementioned methods than using coarser groups of students.

The group popularity ranking method proposed in~\cite{elbadrawy2016domain}
and referred to as {\bf{grp-pop}}, ranks the courses based on how frequently
they were taken by students of the same major and academic level as the
target student. Though this is a simple ranking method, it was shown to be
among the best performing methods proposed by the authors. This is due to the
domain restrictions, where each degree program offers a specific set of
required and elective courses for the students to choose a subset from, and a
pre-requisite structure exists among most of these courses.

\red{\cite{pardos2019connectionist} proposed a similar course2vec model that
  was done independently and in parallel to our proposed work\footnote{An
    earlier version of our paper was published as a technical report at:
    \url{https://goo.gl/HrxVdr}.}. They used a skip-gram neural network
  architecture that takes as input one course, and outputs multiple probability
  distributions over the courses. The approaches that are presented here differ
  from that work because they use a Continuous Bag-Of-Word (CBOW) neural
  network architecture that takes as input multiple courses and outputs one
  probability distribution over the courses for recommendation. Another
  difference is that their model is grade-unaware, while ours is grade-aware,
  which is a main contribution of our work. }

Another model~\cite{backenkohler2018data} that is also parallel and most
relevant to our work also proposed to combine grade prediction with course
recommendation. Our work is different in three aspects.

First, \cite{backenkohler2018data} uses a course dependency graph constructed
using the Mann-Whitney U-test as the course recommendation method. This graph
consists of nodes that represent courses and directed edges between them. A
directed edge going from course A to course B means that the chance of getting
a better grade in B is higher when A is taken before B than when A is not taken
before B. One limitation of this approach is that, for pairs (A, B) of courses
that do not have sufficient data about A not being taken before B, no directed
edge will exist from A to B, despite the fact that there may be sufficient data
about A followed by B, which may imply that A is a pre-requisite for B. Our
proposed representation learning approaches for course recommendation,
described in Section~\ref{sec:method:represent-learn}, on the other hand, are
able to learn all possible orderings for pairs of courses that have sufficient
data. In addition, the course embeddings are learned in a way such that courses
that are taken after a common set of courses are located close in the latent
space, which enables discovering new relationships between previous and
subsequent courses that do not necessarily exist in the data.

Second, we propose a new additional approach for grade-aware course
recommendation, which modifies the course recommendation objective function to
differentiate between good and bad sequences of courses and does not require a
grade prediction method.

\subsection{Course Sequence Discovery and Recommendation}

Though our focus in this paper is to recommend courses for students in their
following term, and not to recommend the whole sequence of courses for all
terms, our proposed models try to learn the sequencing of courses such that
they predict the next-term's good courses based on the previously-taken set of
courses.

\citeN{cucuringu2017rank} utilized several ranking algorithms, e.g., PageRank,
to extract a global ranking of the courses, where the rank here denotes the
order in which the courses are taken by students. The discovered course
sequences were used to infer the hidden dependencies, i.e., informal
prerequisites, between the courses, and to understand how/if course sequences
learned from high- and low-performing students are different from each
other. This technique learns only one global ranking of courses from all
students, which cannot be used for personalized recommendation.

\citeN{xu2016personalized} proposed a course sequence recommendation framework
that aimed to minimize the time-to-graduate, which is based on satisfying the
pre-requisite requirements, course availability during the term, the maximum
number of courses that can be taken during each term, and the degree
requirements. They also proposed to do joint optimization of both graduation
time and GPA by clustering students based on some contextual information, e.g.,
their high school rank and SAT scores, and keeping track of each student's
sequence of taken courses as well as his/her GPA. Then, for a new student,
he/she is assigned to a specific cluster based on their contextual information
and the sequence of courses from that cluster that has the highest GPA estimate
is recommended to him/her. This framework can work well on the more restricted
degree programs that have little variability between the degree plans taken by
students, given that there is enough support for the different degree plans
from past students. However, the more flexible degree programs have much
variability in the degree plans taken by their students, as shown
in~\cite{morsy2019study}. This makes an exact extraction system like the one
above inapplicable for their students, unless there exists a huge dataset that
covers the many different possible sequences with high support.

\subsection{Representation Learning}
\label{sec:rel:rep-learn}

Representation learning has been an invaluable approach in machine learning and
artificial intelligence for learning from different types of data such as text
and graphs. Objects can be represented in a vector space via local or
distributed representations. Under local (or one-hot) representations, each
object is represented by a binary vector, of size equal to the total number of
objects, where only one of the values in the vector is one and all the others
are set to zero. Under distributed representations, each object is represented
by a dense or sparse vector, which can come from hand-engineered features that
is usually sparse and high-dimensional, or a learned representation, called
``embeddings'' in a latent space that preserves the relationships between the
objects, which is usually low-dimensional and more practical than the former.

A widely used approach for learning object embeddings is Singular Value
Decomposition (SVD)~\cite{golub1970singular}. SVD is a traditional low-rank
approximation method that has been used in many fields. In recommendation
systems, a user-item rating matrix is typically decomposed into the user and
item latent factors that uncover the observed ratings in the matrix~\cite[for
eg.]{sarwar2000application,bell2007modeling,paterek2007improving,koren2008factorization}.

Recently, neural networks have gained a lot of interest for learning object
embeddings in different fields, for their ability to handle more complex
relationships than SVD. Some of the early well-known architectures include
Word2vec~\cite{mikolov2013distributed} and Glove~\cite{pennington2014glove},
which were proposed for learning distributed representations for
words~\cite{mikolov2013distributed}. For instance, neural language models for
words, phrases and documents in Natural Language Processing~\cite[for
eg.]{huang2012improving,mikolov2013efficient,le2014distributed,pennington2014glove,mikolov2013distributed}
are now widely used for different tasks, such as machine translation and
sentiment analysis. Similarly, learning embeddings for graphs, such as:
DeepWalk~\cite{perozzi2014deepwalk}, LINE~\cite{tang2015line} and
node2vec~\cite{grover2016node2vec} were shown to have performed well on
different applications, such as: multi-label classification and link
prediction. Moreover, learning embeddings for products in e-commerce and music
playlists in cloud-based music services have been recently proposed for next
basket
recommendation~\cite{chen2012playlist,grbovic2015commerce,wang2015learning}.

\section{Grade-aware Course Recommendation}
\label{sec:method}

\begin{figure}[t]
  \centering
  \includegraphics[width=0.6\textwidth]{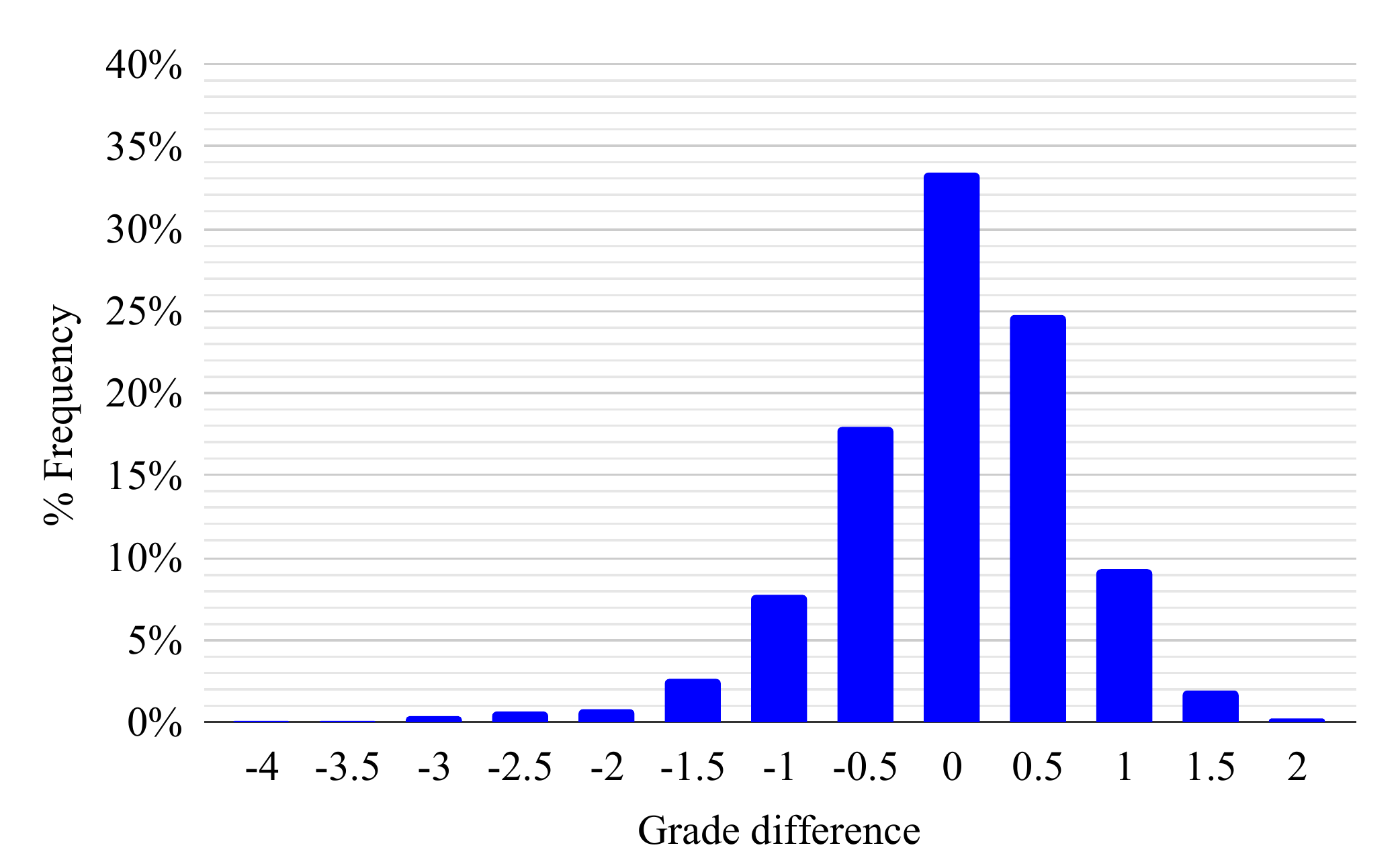}\\
  \caption{\label{fig:grade-dev}Grade difference
    from the student's average previous grade.}

\end{figure}

Undergraduate students often achieve inconsistent grades in the various courses
they take, which may increase or decrease their overall GPA. This is
illustrated in Figure~\ref{fig:grade-dev} that shows the histogram of
differences between each grade obtained by a student over his/her prior average
grade, for the dataset used in our experiments (Table~\ref{tbl:stats}). As we
can see, more than $10\%$ of the grades are a full-letter grade lower, than the
corresponding students' previous average grades\footnote{The letter grading
  system in this dataset has 11 letter grades (A, A-, B+, B, B-, C+, C, C-, D+,
  D, F) that correspond to the numerical grades (4, 3.67, 3.33, 3, 2.67, 2.33,
  2, 1.67, 1.33, 1, 0), with A being the highest grade and F the lowest
  one.}. The poor performance in some of these courses can result in students
having to retake the same courses at a later time, or increase the number of
courses that they will have to take in order to graduate with a desired GPA. As
a result, this will increase the financial cost associated with obtaining a
degree and can incur an opportunity cost by delaying the students' graduation.

For the cases in which a student's performance in a course is a result of
him/her not being well-prepared for it (i.e., is taking the course at the wrong
time in his/her studies), course recommendation methods can be used to
recommend a set of courses for that student that will help: (i) him/her in
completing his/her degree in a timely fashion, and (ii) maintain or improve
his/her GPA. We will refer to the methods that do those simultaneously as {\bf
  grade-aware course recommendation} approaches. Note that the majority of the
existing approaches cannot be used to solve this problem as they ignore the
performance the student is expected to get in the courses that they recommend.

In this work, we propose two different approaches for grade-aware course
recommendation. The first approach (Section~\ref{sec:method:represent-learn})
uses two representation learning approaches that explicitly differentiate
between courses in which the student is expected to perform well in and courses
in which the student is expected not to perform well in. The second approach
(Section~\ref{sec:method:combining}) combines grade prediction methods with
course recommendation methods to improve the final course rankings. The goal of
both approaches is to rank the courses in which the student is expected to
perform well in higher than those in which he/she is expected not to perform
well in.

\subsection{Grade-aware Representation Learning Approaches}
\label{sec:method:represent-learn}

Our first approach for solving the grade-aware course recommendation problem
relies on modifying the way we use the previous students' data to differentiate
between courses which the student is expected to perform well in and courses
which the student is expected not to perform well in. As such, for every
student, we define a course taken by him/her to be \ul{\bf a good (subsequent)
  course} if the student's grade in it is equal to or higher than his/her
average previous grade, otherwise, we define that course to be \ul{\bf a bad
  (subsequent) course}. The goal of our method is to recommend to each student
a set of good courses.

Motivated by the success of representation learning approaches in
recommendation
systems~\cite{koren2008factorization,chen2012playlist,grbovic2015commerce,wang2015learning},
we adapt two widely-used approaches to solve the grade-aware course
recommendation problem. The first approach applies Singular Value Decomposition
linear factorization model on a co-occurrence frequency matrix that
differentiates between good and bad courses
(Section~\ref{sec:method:represent-learn:svd}), while the second one optimizes
an objective function of a neural network log-linear model that differentiates
between good and bad courses (Section~\ref{sec:method:represent-learn:c2v}).

In both approaches, the courses taken by each student are treated as
temporally-ordered sets of courses, and each approach is trained on this data
in order to learn the proper ordering of courses as taken by students. The
course representations learned by these models are then used to create
personalized rankings of courses for students that are designed to include
courses that are relevant to the students' degree programs and will help them
maintain or increase their GPAs.

\subsubsection{{\bf Singular Value Decomposition (SVD)}}
\label{sec:method:represent-learn:svd}

SVD~\cite{golub1970singular} is a traditional low-rank linear model that has
been used in many fields. It factorizes a given matrix $\mathbf{X}$ by finding
a solution to $\mathbf{X} = \mathbf{U} \bm{\Sigma} \mathbf{V}^T$, where the
columns of $\mathbf{U}$ and $\mathbf{V}$ are the left and right singular
vectors, respectively, and $\bm{\Sigma}$ is a diagonal matrix containing the
singular values of $\mathbf{X}$. The $d$ largest singular values, and
corresponding singular vectors from $\mathbf{U}$ and $\mathbf{V}$, is the rank
$d$ approximation of $\mathbf{X}$
($\mathbf{X}_d = \mathbf{U}_d \bm{\Sigma}_d \mathbf{V}_d^T$). This technique is
called truncated SVD.

Since we are interested in learning course ordering as taken by past students,
we apply SVD on a previous-subsequent co-occurrence frequency matrix
$\mathbf{F}$, where $F_{ij}$ is the number of students in the training data
that have taken course $i$ before they took course $j$.

We form two different previous-subsequent co-occurrence frequency matrices, as
follows. Let $n_{ij}^+$ and $n_{ij}^-$ be the number of students who have taken
course $i$ before course $j$, where course $j$ is considered a good course for
the first group and a bad course for the second one, respectively. The two
matrices are:
\begin{enumerate}
\item \freqmatplus: where $F_{ij}^+ = n_{ij}^+$.
\item \freqmatplusminus: where $F_{ij}^{+-} = n_{ij}^+ - n_{ij}^-$.
\end{enumerate}
We scaled the rows of each matrix to $L1$ norm and then applied truncated SVD
on them. The course embeddings are then given by
$\mathbf{U}_d \sqrt{\bm{\Sigma}_d}$ and $\mathbf{V}_d \sqrt{\bm{\Sigma}_d}$ for
the previous and subsequent courses, respectively.

Note that we append a (+), or (+-) as a superscript to the matrix and as a
suffix to the corresponding method's name based on what course information it
utilizes during learning and how it utilizes it. A (+)-based method utilizes
the good course information only and ignores the bad ones, while a (+-)-based
method utilizes both the good and bad course information and differentiates
between them.

\paragraph{{\bf Recommendation.}}

Given the previous and subsequent course embeddings estimated by SVD, course
recommendation is done as follows. Given a student $s$ with his/her
previously-taken set of courses, $c_1, \dots, c_k$, who would like to register
for his/her following term, we compute his/her implicit profile by averaging
over the embeddings of the courses taken by him/her in all previous
terms\footnote{Note that we tried using different window sizes for the number
  of previous terms. Using all previous terms achieved the best results than
  using one, two or three previous terms only.}. We then compute the dot
product between $s$'s profile and the embeddings of each candidate course
$c_t \in C$. Then, we rank the courses in non-increasing order according to
these dot products, and select the top courses as the final recommendations for
$s$.

\subsubsection{{\bf \model}}
\label{sec:method:represent-learn:c2v}

The above SVD model works on pairwise, one-to-one relationships between
previous and subsequent courses. We also model course ordering using a
many-to-one, log-linear model, which is motivated by the recent word2vec
Continuous Bag-Of-Word (CBOW) model~\cite{mikolov2013efficient}. Word2vec works
on sequences of individual words in a given text, where a set of nearby
(context) words (i.e., words within a pre-defined window size) are used to
predict the target word. In our case, the sequences would be the ordered terms
taken by each student, where each term contains a set of courses, and the
previous set of courses would be used to predict future courses for each
student.

\paragraph{Model Architecture.}

\begin{figure}[t]
  \centering
  \includegraphics[width=0.55\textwidth]{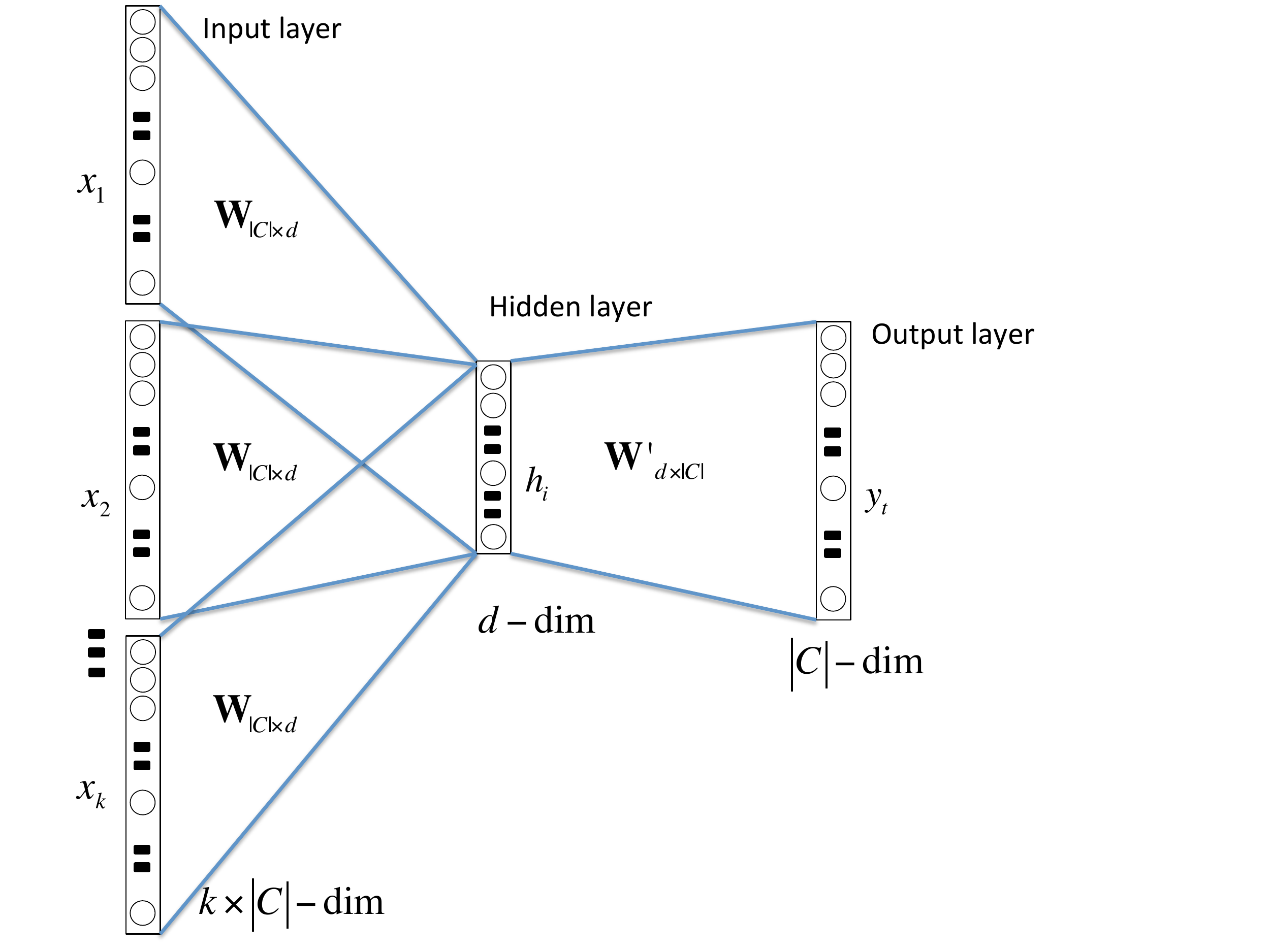}
  \caption{\label{fig:arch}Neural network architecture for \model.}
\end{figure}

We formulate the problem as a maximum likelihood estimation problem. Let
$\term_i = \{c_1, \dots, c_n\}$ be a set of courses taken in some term $i$. A
sequence $\seq_s = \langle \term_1, \dots, \term_m \rangle$ is an ordered list
of $m$ terms as taken by some student $s$, where each term can contain one or
more courses. Let $\inmat \in \R^{|C| \times d}$ be the courses' representations
when they are treated as \emph{previous} courses, and let
$\outmat \in \R^{d \times |C|}$ be their representations when they are treated
as ``subsequent'' courses, where $|C|$ is the number of courses and $d$ is the
number of dimensions in the embedding space. We define the probability of
observing a future course $c_t$ given a set of previously-taken courses
$c_1, \dots, c_k$ using the softmax function, i.e.,
\begin{equation}
  \label{eq:prob}
  Pr(c_t | c_1, \dots, c_k) = y_t = \frac{\exp(\outvec_{c_t}^T \hidden)}{\sum_{j=1}^C
    \exp(\outvec_{c_j}^T \hidden)},
\end{equation}
\noindent where $\hidden$ denotes the aggregated vector of the representations
of the previous courses, where we use the average pooling for aggregation,
i.e.,
\begin{equation}
  \hidden = \frac{1}{k} \inmat^T (\mathbf{x}_1 + \mathbf{x}_2 + \dots +
  \mathbf{x}_k), \nonumber
\end{equation}
\noindent where $\mathbf{x}_i$ is a one-hot encoded vector of size $|C|$ that
has $1$ in the $c_i$'s position and $0$ otherwise. The Architecture for \model\
is shown in Figure~\ref{fig:arch}. Note that one may consider more complex neural
network architectures, which is left for future work.

We propose the two following models:

\begin{enumerate}
\item \textbf{\modelplus}. This model maximizes the log-likelihood of observing
  only the good subsequent courses that are taken by student $s$ in some term
  given his/her previously-taken set of courses. The objective function of
  {\modelplus}\ is thus:
  \begin{equation}
    \label{eq:obj:plus}
  \underset{\inmat, \outmat} {\mbox{maximize}} \sum_{s \in \students} \sum_{\term_i \in \seq_s}
  \Big( \log Pr(\good_{s,i} | \prev_{s, i}) \Big),
\end{equation}
\noindent where: $\students$ is the set of students, $\good_{s,i}$ is the set
of good courses taken by student $s$ at term $i$, and $\prev_{s,i}$ is the set
of courses taken by student $s$ prior to term $i$. Note that $i$ starts from
$2$, since the previous set of courses $\prev_{s,i}$ would be empty for $i=1$.

\item \textbf{\modelplusminus}. This model maximizes the log-likelihood of
  observing good courses and minimizes the log-likelihood of observing bad
  courses given the set of previously-taken courses. The objective function of
  {\modelplusminus}\ is thus:
\begin{equation}
  \label{eq:obj:plusminus}
  \begin{array}{ll}
    \underset{\inmat, \outmat} {\mbox{maximize}} \sum_{s \in \students} \sum_{\term_i \in \seq_s}
    \Big( & \log Pr(\good_{s,i} | \prev_{s, i}) \\
          & - \log Pr(\bad_{s,i} | \prev_{s,i})
            \Big), 
    \end{array}
\end{equation}
\noindent where: $\bad_{s,i}$ is the set of bad courses taken by student $s$ at
term $i$, and the rest of the terms are as defined in Eq.~\ref{eq:obj:plus}.

\end{enumerate}

Note that \modelplus\ is analogous to SVD(+) and \modelplusminus\ is analogous
to SVD(+-) in terms of how they utilize the good and bad courses in the
training set.

\paragraph{Model Optimization.}
\label{sec:cr-methods:optimize}

The objective functions in Eqs.~\ref{eq:obj:plus} and~\ref{eq:obj:plusminus}
can be solved using Stochastic Gradient Descent (SGD), by solving for one
subsequent course at a time.  The computation of gradients in the two equations
requires computing Eq.~\ref{eq:prob} for all courses for the denominator, which
requires knowing whether a course is to be considered a good or a bad
subsequent course for a given context. However, not all the relationships
between every context (previous set of courses) and every subsequent course is
known from the data. Hence, for each context, we only update the subsequent
course vector when the course is known to be a good or bad subsequent course
associated with that context. In the case that some context does not have a
sufficient pre-defined number of subsequent courses with known relationships,
then we randomly sample a few other courses and treat them as bad courses,
similar to the negative sampling approach used in
word2vec~\cite{le2014distributed}.

Note that in {\modelplusminus}, since a course can be seen as both a good and a
bad subsequent course for the same context in the data (for different
students), then, in this case, we randomly choose whether to treat that course
as good or bad each time according to a uniform distribution that is based on
its good and bad frequency in the dataset. In addition, for both {\modelplus}
and {\modelplusminus}, if the frequency between a context and a subsequent
course is less than a pre-defined threshold, e.g., 20, then we randomly choose
whether to update that subsequent course's vector in the denominator each time
it is visited. The code for {\model}\ can be found at: \url{https://goo.gl/uCCqie},
which is built on the original word2vec code that was implemented for the CBOW
model\footnote{Original code is at: \url{https://goo.gl/UvUuMQ}}.

\paragraph{{\bf Recommendation}}
\label{sec:cr-methods:recommend}

Given the previous and subsequent course embeddings estimated by \model, course
recommendation is done as follows. Given a student $s$ with his/her
previously-taken set of courses, $c_1, \dots, c_k$, who would like to register
for his/her following term, we compute the probability
$Pr(c_t | c_1, \dots, c_k)$ for each candidate course $c_t \in C$ according to
Eq.~\ref{eq:prob}. We then rank the courses in non-increasing order according
to their probabilities, and select the top courses as the final recommendations
for $s$. Note that since the denominator in Eq.~\ref{eq:prob} is the same for
all candidate courses, the ranking score for course $c_t$ can be simplified to
the dot product between $\outvec_{c_t}$ and $\hidden$, where $\hidden$
represents the student's implicit profile.

\subsection{Combining Course Recommendation with Grade Prediction}
\label{sec:method:combining}

The second approach that we developed for solving the grade-aware course
recommendation problem relies on using the grades that students are expected to
obtain in future courses to improve the ranking of the courses produced by
course recommendation methods. Our underlying hypothesis behind this approach
is that, a course that both is ranked high by a course recommendation method
and has a high predicted grade should be ranked higher than one that either has
a lower ranking by the recommendation method or is predicted to have a lower
grade in it. This in turn will help improve the final course rankings for
students by taking both scores into account simultaneously.

Let $\gradescore_{s, c}$ be the predicted grade for course $c$ as generated
from some grade prediction model, and let $\recscore_{s, c}$ be the ranking
score for $c$ as generated from some course recommendation method. We combine
both scores to compute the final ranking score for $c$ as follows:
\vspace{-10pt}
\begin{equation}
  \label{eq:combined-approach}
  \textrm{\finalscore}_{s, c} ~= \gradescore_{s, c}^\alpha \times
  (\left| \recscore_{s, c} \right|)^{(1-\alpha)} \times
  \textrm{sign}(\recscore_{s, c}),
\end{equation}
\vspace{-2pt}
\noindent where $\alpha$ is a hyper-parameter in the range $(0, 1)$ that
controls the relative contribution of $\gradescore_{s, c}$ and
$\recscore_{s, c}$ to the overall ranking score, and sign($\recscore_{s, c}$)
denotes the sign of $\recscore_{s, c}$, i.e., $1$ if $\recscore_{s, c}$ is
positive and $-1$ otherwise. Note that both $\gradescore_{s, c}$ and
$\recscore_{s, c}$ are standardized to have zero mean and unit variance.

In this work, we will use the representation learning approaches described in
Section~\ref{sec:method:represent-learn} as the course recommendation
method. We will also use the grade-unaware variations of each of them (see
Section~\ref{sec:eval:baseline}) to compare combining the grade prediction
methods with both recommendation approaches.

To obtain the grade prediction score, we will use \emph{Cumulative
  Knowledge-based Regression Models}~\cite{morsy2017cumulative}, or CKRM for
short. CKRM is a set of grade-prediction methods that learn low-dimensional as
well as textual-based representations for courses that denote the required and
provided knowledge components for each course. It represents a student's
knowledge state as the sum of the provided knowledge component vectors of the
courses taken by them, weighted by their grades in them. CKRM then predicts the
student's grade in a future course as the dot product between their knowledge
state vector and the course's required knowledge component vector. We will
denote the recommendation method that combines CKRM with SVD and \model\ as
{\bf CKRM+SVD} and {\bf CKRM+\model}, respectively.

\section{Experimental Evaluation}
\label{sec:eval}

\subsection{Dataset Description and Preprocessing}
\label{sec:eval:dataset}

The data used in our experiments was obtained from the University of Minnesota,
where it spans a period of 16 years (Fall 2002 to Summer 2017). From that
dataset, we extracted the degree programs that have at least 500 graduated
students until Fall 2012, which accounted for 23 different majors from
different colleges. For each of these degree programs, we extracted all the
students who graduated from this program and extracted the 50 most frequent
courses taken by the students as well as the courses that belonged to frequent
subjects, e.g., CSCI is a subject that belongs to the Computer Science
department at the University. A subject is considered to be frequent if the
average number of courses that belong to that subject over all students is at
least three.  \red{This filtering was made to remove the courses we believe are
  not relevant to the degree program of students.} We also removed any courses
that were taken as pass/fail.

Using the above dataset, we split it into train, validation and test sets as
follows. All courses taken before Spring 2013 were used for training, courses
taken between Spring 2013 and Summer 2014 inclusive were used for validation,
and courses taken afterwards (Fall 2014 to Summer 2017 inclusive) were used for
test purposes.

At the University of Minnesota, the letter grading system has 11 letter grades
(A, A-, B+, B, B-, C+, C, C-, D+, D, F) that correspond to the numerical grades
(4, 3.667, 3.333, 3, 2.667, 2.333, 2, 1.667, 1.333, 1, 0). For each (context,
subsequent) pair in the training, validation, and test set, where the context
represents the previously-taken set of courses by a student, the context
contained only the courses taken by the student with grades higher than the D+
letter grade. The statistics of the 23 degree programs are shown in
Table~\ref{tbl:stats}.

\begin{table}[h]
  \caption{Dataset statistics.}
  \centering
  \begin{scriptsize}
    \begin{tabular}{lrrr}
        \toprule
        \multicolumn{1}{c}{Major} & \multicolumn{1}{c}{\# Students} &
                                                                     \multicolumn{1}{c}{\#
                                                                      Courses}
      & \multicolumn{1}{c}{\# Grades} \\
        \midrule
        Accounting (ACCT) & 661 & 55 & 7,614 \\ 
        Aerospace Engr. (AEM) & 866 & 72 & 13,280 \\ 
        Biology (BIOL) & 1,927 & 113 & 15,590 \\ 
        Biology, Soc. \& Envir. (BSE) & 1,231 & 56 & 9,389 \\ 
        Biomedical Engr. (BME) & 1,002 & 64 & 13,808 \\ 
        Chemical Engr. (CHEN) & 1,045 & 82 & 10,219 \\ 
        Chemistry (CHEM) & 765 & 78 & 7,814 \\ 
        Civil Engr. (CIVE) & 1,160 & 74 & 15,992 \\ 
        Communication Studies (COMM) & 2,547 & 90 & 17,135 \\ 
        Computer Science \& Engr. (CSE) & 1,790 & 98 & 13,520 \\ 
        Electrical Engr. (ECE) & 1,197 & 84 & 12,781 \\ 
        Elementary Education (ELEM) & 1,283 & 60 & 15,303 \\ 
        English (ENGL) & 1,790 & 113 & 12,451 \\ 
        Finance (FIN) & 1,326 & 55 & 12,150 \\ 
        Genetics, Cell Biol. \& Devel. (GCD) & 843 & 92 & 9,726 \\ 
        Journalism (JOUR) & 2,043 & 91 & 23,549 \\ 
        Kinesiology (KIN) & 1,499 & 161 & 23,451 \\ 
        Marketing (MKTG) & 2,077 & 51 & 13,084 \\ 
        Mechanical Engr. (MECH) & 1,501 & 79 & 25,608 \\ 
        Nursing (NURS) & 1,501 & 88 & 18,239 \\ 
        Nutrition (NUTR) & 940 & 71 & 12,400 \\ 
        Political Science (POL) & 1,855 & 111 & 13,904 \\ 
        Psychology (PSY) & 3,047 & 100 & 25,299 \\ 
      \bottomrule
    \end{tabular}
    \end{scriptsize}
    \label{tbl:stats}
\end{table}

\subsection{Baseline and Competing Methods}
\label{sec:eval:baseline}

We compare the performance of the proposed representation learning approaches
against competing approaches for grade-aware course recommendation, which are
defined as follows:
\begin{itemize}
\item \textbf{Grp-pop(+-)}: We modify the group popularity ranking method
  developed in~\citeN{elbadrawy2016domain} and explained in
  Section~\ref{sec:related} to solve the grade-aware course recommendation. For
  each course $c$, let $n_c^+$ and $n_c^-$ be the number of students that have
  the same major and academic level as the target student $s$, where $c$ was
  considered a good subsequent course for the first group and a bad one for the
  second group. We can differentiate between good and bad subsequent courses
  using the following ranking score (which is similar to the (+-)-based
  approaches):
  \begin{equation}
    \label{eq:grppopplusminus}
    \textrm{\finalscore}_{s, c} ~= n_c^+  - n_c^-.
  \end{equation}

\item \textbf{Grp-pop(+)}: Here, the group popularity ranking method considers
  only the good subsequent courses, similar to SVD(+) and
  \modelplus. Specifically, the ranking score is computed as:
  \begin{equation}
    \textrm{\finalscore}_c ~= n_c^+, \nonumber
  \end{equation}
  \noindent where $n_c^+$ is as defined in Eq.~\ref{eq:grppopplusminus}.

\item {\bf Course dependency graph}: This is the course recommendation method
  utilized in~\cite{backenkohler2018data} (see
  Section~\ref{sec:related:course-rec}).
\end{itemize}

We also compare the performance of the representation learning approaches for
both grade-aware and grade-unaware course recommendation. The grade-unaware
representation learning approaches are defined as follows:
\begin{itemize}
\item {\bf SVD(++)}: Here, SVD is applied on the previous-subsequent
  co-occurrence frequency matrix: \freqmatplusplus: where
  $F_{ij}^{++} = n_{ij}^+ + n_{ij}^-$.
\item {\bf \modelplusplus}. This model maximizes the log-likelihood of
  observing all courses taken by student $s$ in some term given the set of
  previously-taken courses, regardless of the subsequent course being a good or
  a bad one. This can be written as:
\begin{equation}
  \label{eq:c2v++}
  \underset{\inmat, \outmat} {\mbox{maximize}} \sum_{s \in \students} \sum_{\term_i \in \seq_s}
  \Big( \log Pr(\mathcal{C}_{s,i} | \prev_{s, i}) \Big), \nonumber
\end{equation}
\noindent where: $\mathcal{C}_{s,i}$ is the set of courses taken by student $s$
at term $i$, and the rest of the terms are as defined in Eq.~\ref{eq:obj:plus}.
\end{itemize}

\noindent Note that, here we append a (++) suffix to the grade-unaware
variation of the method's name since it utilizes all the course information
without differentiating between good and bad courses.

\subsection{Evaluation Methodology and Metrics}
\label{sec:eval:metrics}

Previous course recommendation methods used the recall metric to evaluate the
performance of their methods. The goal of the proposed grade-aware course
recommendation methods is to recommend to the student courses which he/she is
expected to perform well in and not recommend courses which he/she is expected
not to perform well in. As a result, we cannot use the recall metric as is, and
instead, we use three variations of it that differentiate between good and bad
courses. The first, \emph{\goodrecall}, measures the fraction of the actual
good courses that are retrieved. The second, \emph{\badrecall}, measures the
fraction of the actual bad courses that are retrieved. The third,
\emph{\diffrecall}, measures the overall performance of the recommendation
method in ranking the good courses higher than the bad ones.

The first two metrics are computed as the average of the student-term-specific
corresponding recalls. In particular, for a student $s$ and a target term $t$,
the first two recall metrics for that ($s$, $t$) tuple are computed as:
\begin{enumerate}
\item $\textrm{\goodrecall}_{(s,t)} = \frac{\left| G_{s,n_{(s,t)}} \right|}{n^g_{(s,t)}}$.
\item $\textrm{\badrecall}_{(s,t)} = \frac{\left| B_{s,n_{(s,t)}}
    \right|}{n^b_{(s,t)}}$.
\end{enumerate}
\noindent $G_{s,n_{(s,t)}}$ and $B_{s,n_{(s,t)}}$ denote the set of good and
bad courses, respectively, that were taken by $s$ in $t$ and exist in his/her
list of $n_{(s,t)}$ recommended courses, $n_{(s, t)}$ is the actual number of
courses taken by $s$ in $t$, and $n^g_{(s,t)}$ and $n^b_{(s,t)}$ are the actual
number of good and bad courses taken by $s$ in $t$, respectively. Since our
goal is to recommend good courses only, we consider a method to perform well
when it achieves a high \goodrecall\ and a low
\badrecall.

\diffrecall\ is computed as the difference between \goodrecall\ and \badrecall,
i.e.,
\begin{itemize}
\item [3.] \textrm{\diffrecall} = \goodrecall\ ~- ~\badrecall.
\end{itemize}
\noindent \diffrecall\ is thus a signed measure that assesses both the degree
and direction to which a recommendation method is able to rank the actual good
courses higher than the bad ones in its recommended list of courses for each
student, so the higher the \diffrecall\ value, the better the recommendation
method is.

To further analyze the differences in the ranking results of the proposed
approaches, we also computed the following two metrics:
\begin {itemize}
\item {\bf Percentage GPA increase/decrease:} Let $\textrm{\curgoodgpa}_s$ and
  $\textrm{\curbadgpa}_s$ be the current GPA achieved by student $s$ on the
  good and bad courses recommended by some recommendation method, respectively,
  and let $\textrm{\prevgpa}_s$ be his/her GPA prior to that term. Then, the
  percentage GPA increase and decrease are computed as:
  \begin{equation}
    \textrm{\% GPA increase} = \frac{\textrm{\curgoodgpa}_s -
      \textrm{\prevgpa}_s}{\textrm{\prevgpa}_s} \times 100.0. \nonumber
  \end{equation}
  \begin{equation}
    \textrm{\% GPA decrease} = \frac{\textrm{\prevgpa}_s -
      \textrm{\curbadgpa}_s}{\textrm{\prevgpa}_s} \times 100.0. \nonumber
  \end{equation}
\item {\bf Coverage for good/bad terms:} The number of terms where some
  recommendation method recommends good (or bad) subsequent courses to will be
  referred to as its coverage for good (or bad) terms. The higher the coverage
  for good terms by some method, the more students who will get good
  recommendations that will maintain or improve their overall GPA. On the other
  hand, the lower the coverage for bad terms, the less students who will get
  bad recommendations that will decrease their overall GPA.
\end{itemize}
We compute the above two metrics for the terms on which the recommendation
method recommends at least one of the actual courses taken in that term. For
each method, the percentage GPA increase and decrease as well as the coverage
for good and bad terms are computed as the average of the individual
scores. Since we would like to recommend courses that optimize the student's
GPA, the higher the GPA percentage increase and the coverage for good terms and
the lower the GPA percentage decrease and the coverage for bad terms that a
method achieves, the better the method is.

Note that, a recommendation is only done for students who have taken at least
three previous courses. For each ($s$, $t$) tuple, the recommended list of
courses using any method are selected from the list of courses that are being
offered at term $t$ only, and that were not already taken by $s$ with an
associated grade that is either: (i) $\ge \textrm{C+}$, or, (ii)
$\ge \mu_s - 1.0$, where $\mu_s$ is the average previous grade achieved by
$s$. Therefore, we only allow recommending repeated courses in the case that
the student has achieved a low grade in it such that the course's credits do
not add to the earned credits, or when they a achieve bad grade in them
relative to their grades in previous terms. This filtering technique
significantly improved the performance of all the baseline and proposed
methods.

\subsection{Model Selection}
\label{sec:eval:model-select}

We did an extensive search in the parameter space for model selection. The
parameters in the SVD-based models is the number of latent dimensions
($d$). The parameters in the {\model}-based models are: the number of latent
dimensions ($d$), and the minimum number of subsequent courses ($samples$), in
the denominator of Eq.~\ref{eq:prob} that are used during the SGD process of
learning the model.  We experimented with the parameter $d$ in the range
$[10-30]$ with a step of $5$, with the minimum number of $samples$ with the
values $\{3, 5\}$ , and with the parameter $\alpha$ in
Eq.~\ref{eq:combined-approach} in the range $[0.1-0.9]$ with a step of 0.2.

The training set was used for learning the distributed representations of the
courses, whereas the validation set was used to select the best performing
parameters in terms of the highest \diffrecall. 

\section{Results}
\label{sec:res}

We evaluate the effectiveness of the proposed grade-aware course recommendation
methods in order to answer the following questions:
\begin{itemize}
\item[RQ1.] How do the SVD- and Course2vec-based approaches for course
  recommendation compare to each other?
\item [RQ2.] How do the combination of grade prediction with representation
  learning approaches compare to each other?
\item [RQ3.] How do the two proposed approaches for solving grade-aware course
  recommendation compare to each other?
\item [RQ4.] How do the proposed approaches compare to competing approaches for
  grade-aware course recommendation?
\item [RQ5.] What are the benefits of grade-aware course recommendation over
  grade-unaware course recommendation?
\item [RQ6.] \red{How does the recommendation accuracy vary across different
    majors and student sub-groups?}
\item [RQ7.] \red{What are the characteristics of the recommended courses, in
    terms of course difficulty and popularity?}
\end{itemize}

\subsection{Comparison of the Representation Learning Approaches for
  Grade-aware Course Recommendation}
\label{sec:res:grade-rep-learn}

Table~\ref{tbl:grade-aware-rep-learn} shows the prediction performance of the
two proposed representation learning approaches for grade-aware course
recommendation. The results show that SVD(+) achieves the best \goodrecall,
while SVD(+-) achieves the best \diffrecall. \modelplusminus\ achieves the best
\badrecall, which is comparable to SVD(+-).

By comparing the corresponding SVD and \model\ approaches, we see that SVD
outperforms \model\ in almost all cases. We believe this is caused by the fact
that there is a limited number of positive training data for \model, since only
the good courses are used as positive examples for learning the models. This is
supported by the comparable prediction performance of the (++)-based approaches
that use all the available training data as positive examples, which are shown
in Table~\ref{tbl:grade-aware-unaware-rep-learn}.

By comparing the (+)- and (+-)-based methods, we see that, the (+-)-based model
achieves a worse \goodrecall, but a much better \badrecall. For instance,
SVD(+-) achieves a $15\%$ decrease in \goodrecall\ and a $45\%$ decrease in
\badrecall\ over SVD(+). This is expected, since utilizing the bad course
information gives the models more power to learn to rank these courses low, but
it also adds some noise, since different students with the same or similar
previous set of courses can achieve different outcomes on the same courses.

\begin{table}[t]
  \centering
  \caption{Prediction performance of the proposed representation learning based
    approaches for grade-aware course recommendation.}
   \begin{scriptsize}
     \begin{threeparttable}
       \begin{tabular}{lrrrr}
         \toprule
         \multirow{2}{*}{Metric} & \multicolumn{2}{c}{SVD} & \multicolumn{2}{c}{\model} \\
                                 & \multicolumn{1}{c}{(+)} &
                                                             \multicolumn{1}{c}{(+-)} & \multicolumn{1}{c}{(+)} & \multicolumn{1}{c}{(+-)} \\
         \midrule

         \goodrecall & \ul{0.468} & 0.396  & 0.448 & 0.351 \\
         \badrecall & 0.372 & 0.206 & 0.404 & \ul{0.202} \\
         \diffrecall & 0.096 & \ul{0.190} & 0.044 & 0.149 \\

         \bottomrule
       \end{tabular}
     \end{threeparttable}
     \label{tbl:grade-aware-rep-learn}
     \begin{minipage}{0.9\textwidth}
       The underlined entries denote the results with the best performance for
       each metric.
     \end{minipage}
   \end{scriptsize}
 \end{table}

\subsection{Comparison of the Grade-aware Recommendation Approaches Combining
  Grade Prediction with Course Recommendation}

Table~\ref{tbl:ckrm-combined} shows the prediction performance of the
grade-aware recommendation approaches that combine CKRM with the grade-aware
and grade-unaware representation learning methods. The results show that
CKRM+SVD(++) achieves the best \goodrecall, while CKRM+Course2vec(+-) achieves
the best \badrecall. Overall, CKRM+SVD(+-) achieves the best
\diffrecall. Combining CKRM with the grade-unaware, i.e., (++)-based,
approaches helped in differentiating between good and bad courses, by achieving
a high \diffrecall\ of 0.158 and 0.142 for SVD and \model,
respectively. However, despite these performance improvements, the combinations
that use the grade-aware recommendation methods do better. For instance,
CKRM+SVD(+) outperforms CKRM+SVD(++) by $15\%$ in terms of \diffrecall.

The results also show that the SVD-based (+)- and (+-)-based approaches
outperform their \model\ counterparts in terms of \diffrecall, similar to the
results of SVD and Course2vec alone
(Section~\ref{sec:res:grade-rep-learn}). Unlike the difference in the
performance of SVD(+) vs SVD(+-), CKRM+SVD(+) achieves a similar \diffrecall\
to that achieved by CKRM+SVD(+-) (and the same holds for the Course2vec-based
approaches). The difference is that CKRM+SVD(+) achieves higher \goodrecall\
and \badrecall\ than CKRM+SVD(+-).

\subsection{Comparison of the Proposed Approaches for Grade-aware Course
  Recommendation}

Comparing each of the SVD- and Course2vec-based approaches with and without
CKRM (shown in Tables~\ref{tbl:grade-aware-rep-learn}
and~\ref{tbl:ckrm-combined}), we see that combining CKRM with the (+)-based
approaches significantly improved their performance with $95\%$ and $245\%$
increase in \diffrecall\ for SVD and \model, respectively. On the other hand,
combining CKRM with the (+-)-based approaches achieves comparable performance
to using the corresponding (+-)-based approach alone.

\begin{figure}[t]
  \centering \subfloat[Percentage GPA increase and
  decrease\label{fig:svd_gpa_increase_decrease}]{
     \includegraphics[width=0.48\textwidth]{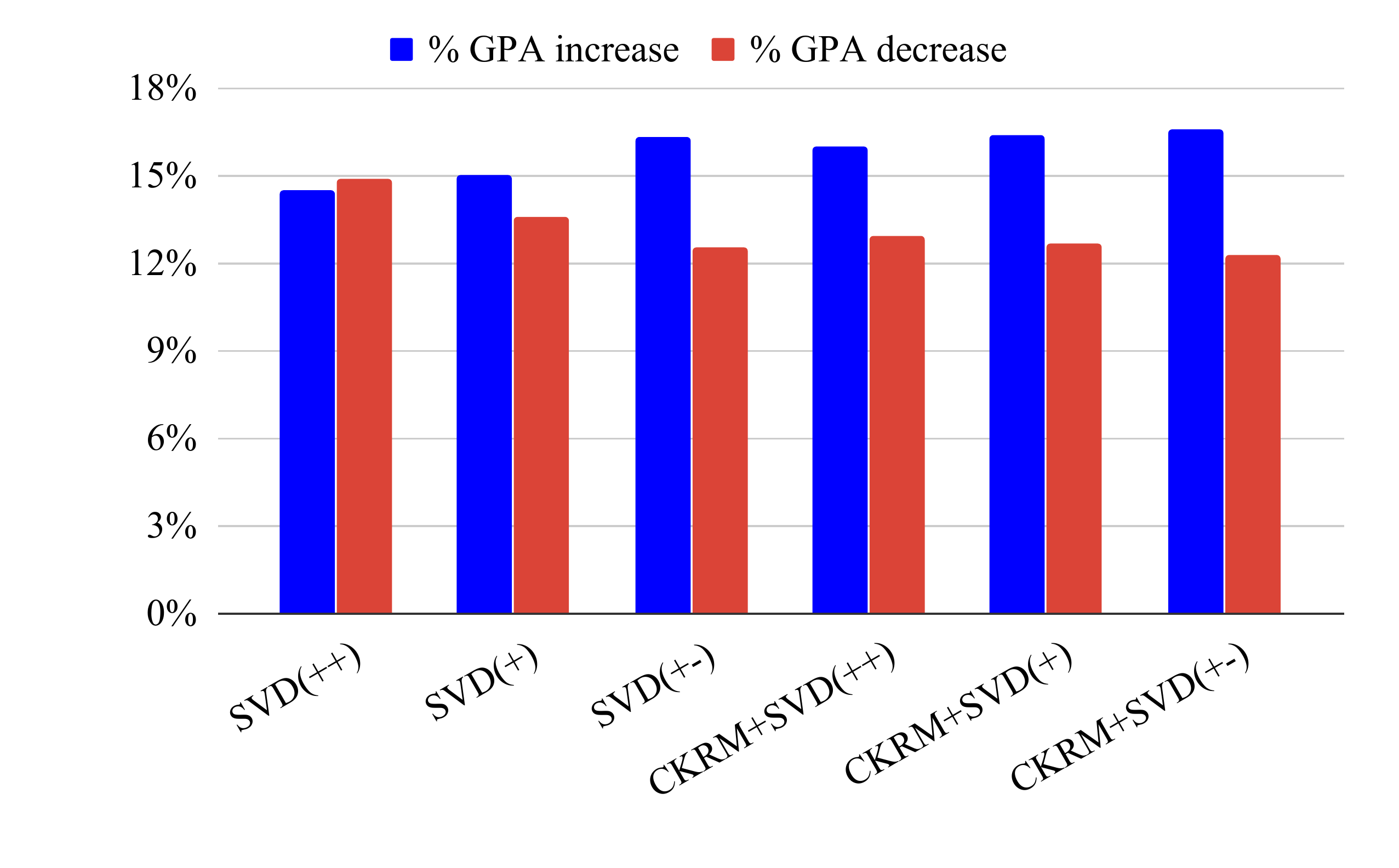} }
    \hfill
    \subfloat[Coverage for good and bad terms\label{fig:svd_n_good_bad_terms}]{
      \includegraphics[width=0.48\textwidth]{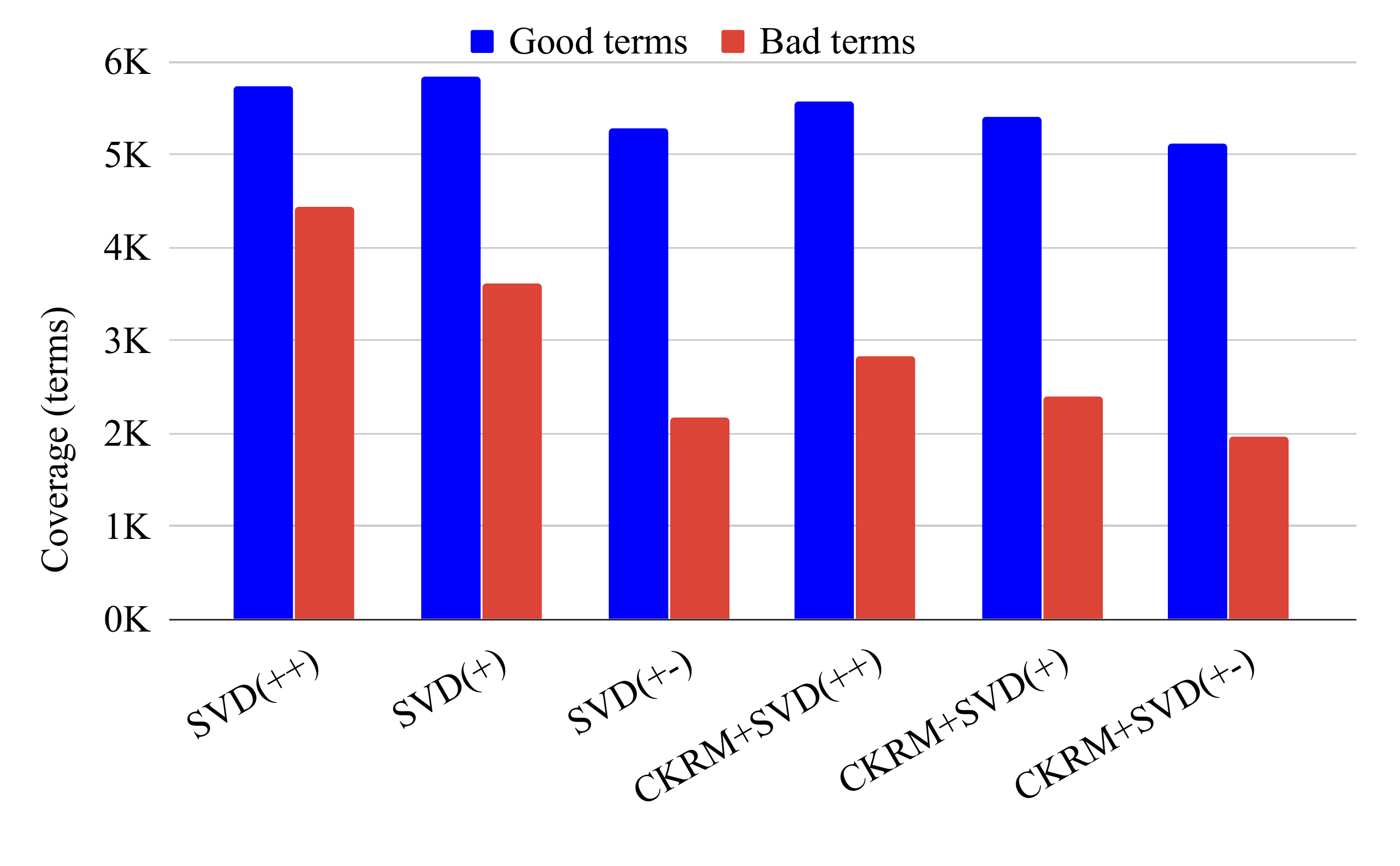} }
    \caption{\label{fig:svd_gpa_increase_decrease_term_coverage}Performance of
      the different SVD-based methods with and without CKRM (refer to
      Section~\ref{sec:eval:metrics} for the metrics definitions).}
\end{figure}

By further analyzing these ranking results,
Figure~\ref{fig:svd_gpa_increase_decrease_term_coverage} shows the percentage
GPA increase and decrease as well as the coverage for good and bad terms for
each SVD-based method with and without CKRM\footnote{The results of the
  {\model}-based methods are similar, and are thus omitted.}. CKRM+SVD(+)
outperforms SVD(+) in all but one metric, which is coverage for good terms,
where it achieves slightly worse performance than SVD(+). On the other hand,
CKRM+SVD(+-) has comparable performance to SVD(+-), which is analogous to their
recall metrics results.

\begin{table}[t]
  \centering
  \caption{Prediction performance of combining CKRM with the representation
    learning based approaches for grade-aware course recommendation methods.}
  \begin{scriptsize}
    \begin{threeparttable}
      \begin{tabular}{lrrrrrr}
        \toprule
        \multirow{2}{*}{Metric} & \multicolumn{3}{c}{CKRM + SVD} &
                                                                    \multicolumn{3}{c}{CKRM + \model} \\
                                & \multicolumn{1}{c}{(++)} &
                                                             \multicolumn{1}{c}{(+)} & \multicolumn{1}{c}{(+-)} & \multicolumn{1}{c}{(++)} & \multicolumn{1}{c}{(+)} & \multicolumn{1}{c}{(+-)} \\
        \midrule

        \goodrecall & \ul{0.438} & 0.417 & 0.385 & 0.411 & 0.417 & 0.338 \\
        \badrecall &  0.279 & 0.230 & 0.189 & 0.269 & 0.264 & \ul{0.183} \\
        \diffrecall & 0.158 & 0.187 & \ul{0.197} & 0.142 & 0.152 & 0.155 \\

        \bottomrule
      \end{tabular}
    \end{threeparttable}
    \label{tbl:ckrm-combined}
    \begin{minipage}{0.9\textwidth}
      The underlined entries denote the results with the best performance for
      each metric.
    \end{minipage}
  \end{scriptsize}
\end{table}

\subsection{Representation Learning vs Competing Approaches for
  Grade-aware Course Recommendation}

Table~\ref{tbl:grade-aware-rep-learn-competing} shows the prediction
performance of the representation learning and competing approaches for
grade-aware course recommendation. Grp-pop(+-) achieves the best \diffrecall\
among the three competing (baseline) approaches. The results also show that
SVD(+) achieves the best \goodrecall, while grp-pop(+-) achieves the best
\badrecall. Overall, SVD(+-) achieves the best \diffrecall.

\subsection{Grade-aware vs Grade-unaware Representation Learning Approaches}

Table~\ref{tbl:grade-aware-unaware-rep-learn} shows the performance prediction
of the representation learning approaches for grade-aware, i.e., (+)- and
(+-)-based approaches, and grade-unaware, i.e., (++)-based approach, course
recommendation. Each of SVD(+) and \modelplus\ achieves a \goodrecall\ that is
comparable to or better than that achieved by its corresponding (++)-based
approach. In addition, both the (+)- and (+-)-based methods achieve much better
(lower) \badrecall. For instance, SVD(+) and SVD(+-) achieve $0.372$ and
$0.206$ \badrecall, respectively, resulting in $26\%$ and $59\%$ improvement
over SVD(++), respectively.

\red{By comparing the (++)-, (+)-, and (+-)-based approaches in terms of
  Recall(diff), we can see that the (++)-based approaches achieve negative
  recall values which indicates that they recommend more bad courses than they
  recommend good ones. The (+)-based approaches do slightly better, while the
  (+-)-based approaches achieve the highest Recall(diff)}. This is expected,
since the (++)-based methods treat both types of subsequent courses equally
during their learning, and so they recommend both types in an equal
manner. This shows that differentiating between good and bad courses in any
course recommendation method is very helpful for ranking the good courses
higher than the bad ones, which will help the student maintain or improve their
overall GPA.

In terms of percentage GPA increase and decrease (shown in
Figure~\ref{fig:svd_gpa_increase_decrease_term_coverage}), SVD(+-) outperforms
SVD(++) by $2\%$ in percentage GPA increase and $2.5\%$ in percentage GPA
decrease. Moreover, SVD(+-) achieves $\sim 62\%$ less coverage for the bad
terms than SVD(++), while it achieves $\sim 10\%$ less coverage for the good
terms.

\begin{table}[t]
  \begin{center}
  \caption{Prediction performance of the representation learning based vs
    competing approaches for grade-aware course recommendation.}
  \begin{scriptsize}
    \begin{threeparttable}
      \begin{tabular}{lccccccc}
        \toprule

         \multicolumn{1}{c}{Metric} & \multicolumn{1}{c}{Dependency} &
                                                                       \multicolumn{1}{c}{Grp-pop}
        & \multicolumn{1}{c}{Grp-pop} & \multicolumn{1}{c}{SVD} &
                                                                               \multicolumn{1}{c}{SVD}
        & \multicolumn{1}{c}{\model} & \multicolumn{1}{c}{\model} \\
        & \multicolumn{1}{c}{Graph} & (+) & (+-) & (+) & (+-) & (+) & (+-) \\
        \midrule
        Recall(good) & 0.382 & 0.425 & 0.367 & \ul{0.468} & 0.396 & 0.448 &
                                                                            0.351
      \\
        Recall(bad) & 0.260 & 0.343 & \ul{0.188} & 0.372 & 0.206 & 0.404 &
                                                                           0.202
      \\
        Recall(diff) & 0.122 & 0.082 & 0.179 & 0.096 & \ul{0.190} & 0.044 & 0.149 \\
          \bottomrule
      \end{tabular}
    \end{threeparttable}
    \label{tbl:grade-aware-rep-learn-competing}
    \begin{minipage}{0.9\textwidth}
      The underlined entries denote the results with the best performance for
      each metric.
    \end{minipage}
  \end{scriptsize}
  \end{center}
\end{table}

\begin{table}[t]
  \begin{center}
    \caption{Prediction performance of the representation learning based
      approaches for grade-aware and grade-unaware course recommendation.}
  \begin{scriptsize}
    \begin{threeparttable}
      \begin{tabular}{lcccccc}
        \toprule

        \multicolumn{1}{c}{Metric} & \multicolumn{1}{c}{SVD} &
                                                                    \multicolumn{1}{c}{\model}
        & \multicolumn{1}{c}{SVD} & \multicolumn{1}{c}{\model} &
                                                                         \multicolumn{1}{c}{SVD}
        & \multicolumn{1}{c}{\model} \\
        & (++) & (++) & (+) & (+) & (+-) & (+-) \\
        \midrule
        Recall(good) & 0.453 & 0.455 & \ul{0.468} & 0.448 & 0.396 & 0.351 \\
        Recall(bad) & 0.502 & 0.493 & 0.372 & 0.404 & 0.206 & \ul{0.202} \\
        Recall(diff) & -0.048 & -0.038 & 0.096 & 0.044 & \ul{0.190} & 0.149 \\
        \bottomrule
      \end{tabular}
    \end{threeparttable}
    \label{tbl:grade-aware-unaware-rep-learn}
    \begin{minipage}{0.9\textwidth}
      The underlined entries denote the results with the best performance for
      each metric.
    \end{minipage}
  \end{scriptsize}
  \end{center}
\end{table}

\red{\subsection{Analysis of Recommendation Accuracy}}

\red{Our discussion so far focused on analyzing the performance of the
  different methods by looking at metrics that are aggregated across the
  different majors. However, given that the structure of the degree programs of
  different majors is sometimes quite different, and that different student
  groups can exhibit different characteristics, an important question that
  arise is how the different methods perform across the individual degree
  programs and different student groups and if there are methods that
  consistently perform well across majors as well as across student groups. In
  this section, we analyze the recommendations done by one of our best
  performing models, CKRM+SVD(+-), against the best performing baseline, i.e.,
  \grppopplusminus, in terms of \diffrecall, across these degree programs and
  student groups (RQ6).}

  \red{\subsubsection{Analysis on Different Majors}}
  \label{sec:res:analysis:majors}

\begin{figure}[ht]
  \centering \subfloat[Per-major recommendation accuracy of \grppopplusminus\
  and SVD(+-).\label{fig:recall-diff-per-major}]{
     \includegraphics[width=0.95\textwidth]{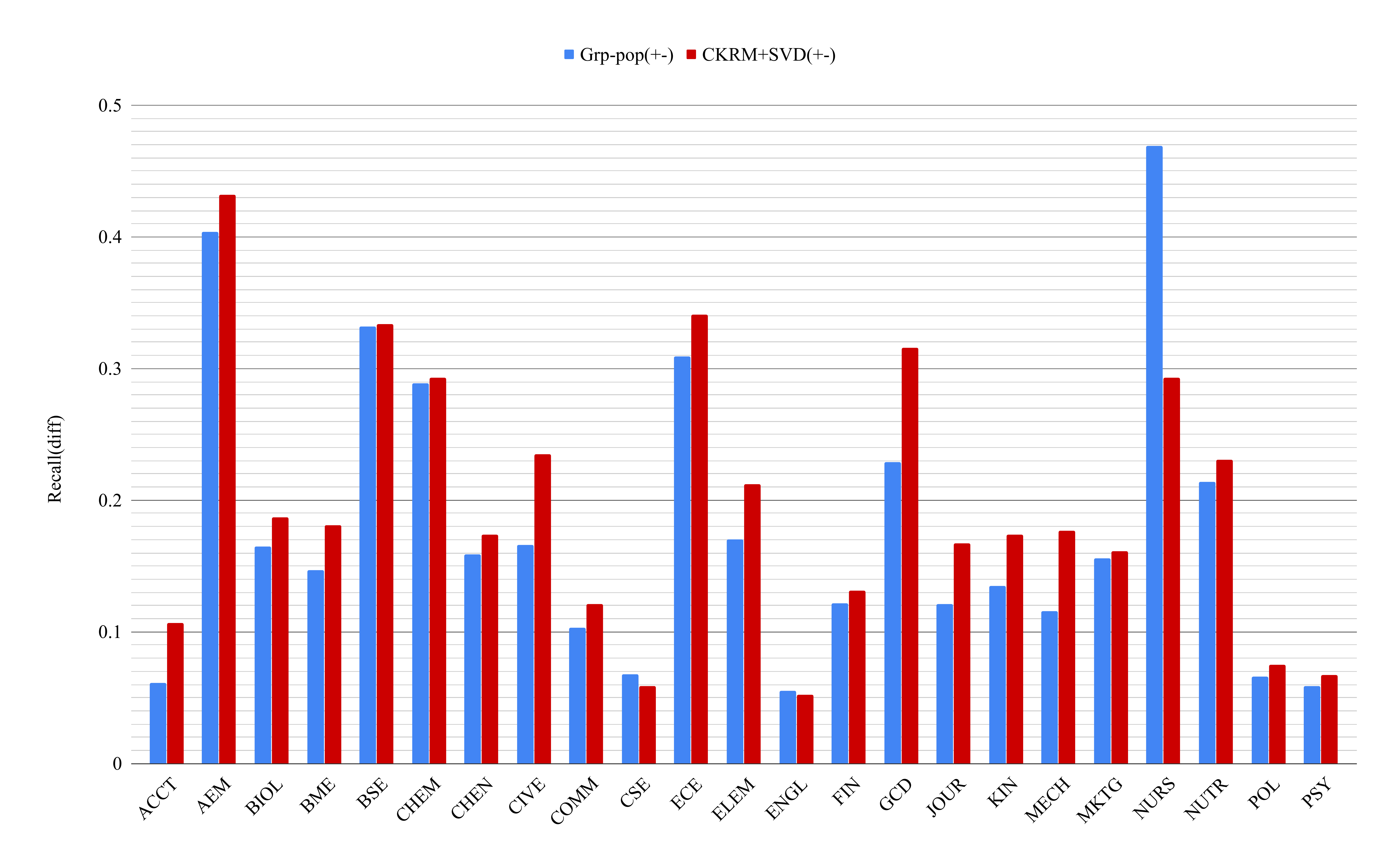} } \\
   \hfill
    \subfloat[Pairwise \% common courses per major.\label{fig:common-courses}]{
      \includegraphics[width=0.48\textwidth]{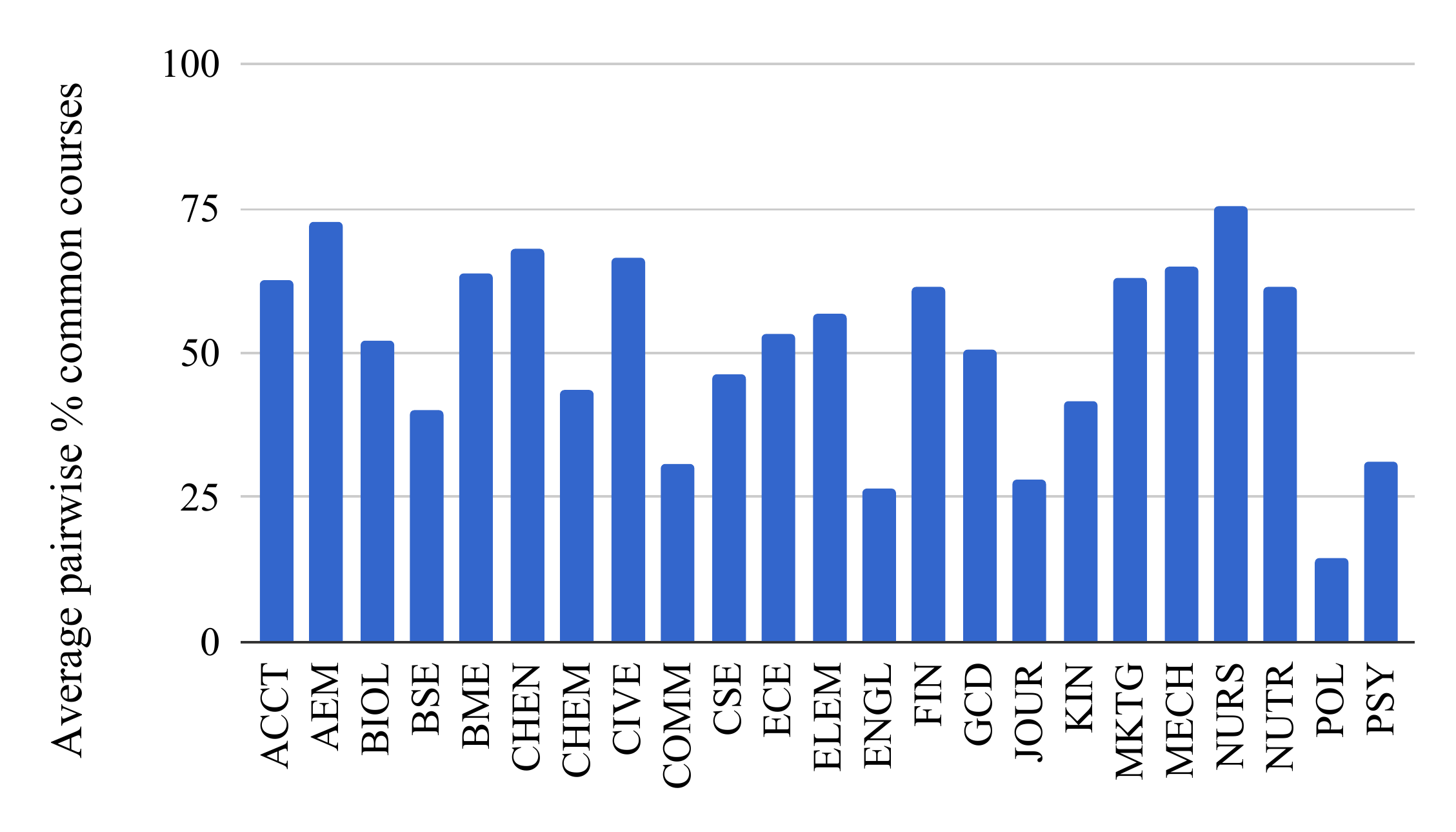} }
    \hfill
    \subfloat[Pairwise degree similarity per major.\label{fig:deg-sim}]{
      \includegraphics[width=0.48\textwidth]{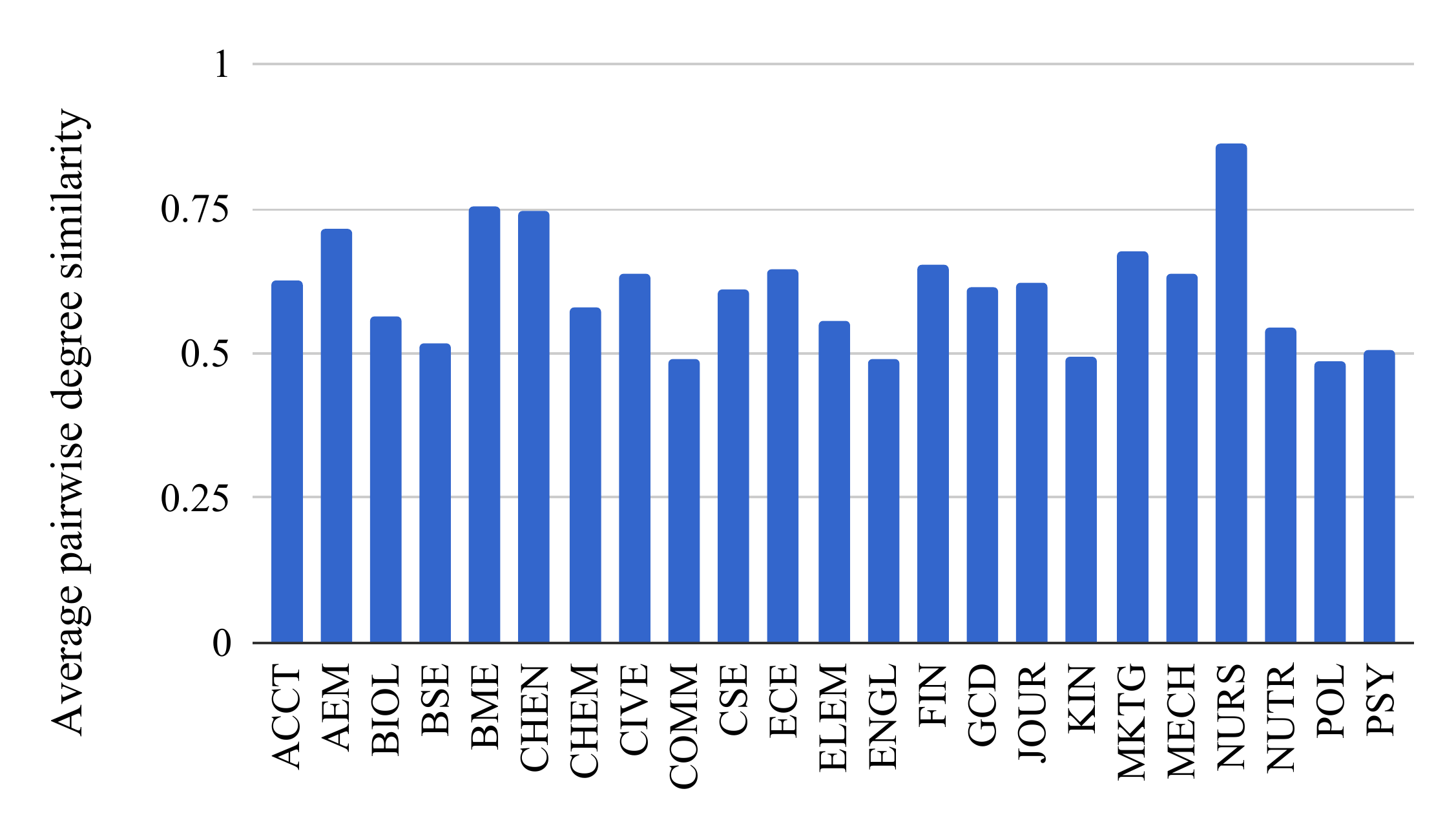} }
    \caption{\label{fig:per-major-analysis}Per-major recommendation accuracy
      and the characteristics of the students' degrees.}
  \end{figure}

\red{Table~\ref{fig:per-major-analysis} shows the recommendation accuracy, in
  terms of Recall(diff), across the 23 majors, by both \grppopplusminus\ and
  CKRM+SVD(+-) (Fig~\ref{fig:recall-diff-per-major}). First, we can see that
  there is a huge variation in the recall values across the majors, ranging
  from 0.05 to $\sim$0.5. Second, we see that CKRM+SVD(+-) consistently
  outperforms \grppopplusminus, except for the nursing major. To further look
  into why this happens, we investigated some of the characteristics of the
  students' degree sequences. For each major, we computed the pairwise
  percentage of common courses among students who belong to that major, which
  is shown in Figure~\ref{fig:common-courses}. In addition, we computed the
  similarity in the sequencing, i.e., ordering, of the common courses between
  each pair of students, which is shown in Figure~\ref{fig:deg-sim}. For
  computing the pairwise degree similarity, we utilized the formula proposed
  in~\cite{morsy2019study}, which computes the degree similarity between a pair
  of degree plans $d_1$ and $d_2$ as:} \red{\begin{equation}
  \label{eq:seq-sim}
  \textrm{sim}(d_1, d_2) = \frac{\sum_{(x,y) \in |C_1 \cap C_2|} T(t_{1,x}-t_{1,y},
    t_{2,x}-t_{2,y})}{|C_1 \cap C_2|},
\end{equation}}
\noindent \red{where $C_i$ is the set of courses taken in degree $i$, and $t_{i, x}$ is
the time, i.e., term number, that course $x$ was taken in $d_i$, e.g., the
first term is numbered 1, the second is numbered 2 and so forth. Function
$T(dt_1, dt_2)$ is defined as:}
\red{\begin{equation}
  T(dt_1, dt_2) =
  \begin{cases}
    1, & \text{if } dt_1 = dt_2 = 0 \\
    \exp \Big({-\lambda (\left| dt_1 - dt_2 \right|)} \Big), & \text{if } dt_1 \times dt_2 \ge 1 \\
    0,                               & \text{otherwise}.
  \end{cases}
\end{equation}}
\noindent \red{where $\lambda$ is an exponential decay constant. Function $T$
assigns a value of $1$ for pairs of courses taken concurrently, i.e., during
the same term, in both plans, and assigns a value of $0$ for pairs of courses
that are either: (i) taken in reversed order in both plans, or (ii) taken
concurrently in one plan and sequentially in the other. For pairs of courses
taken in the same order, it assigns a positive value that decays exponentially
with $\left| dt_1 - dt_2 \right|$.}

\red{We found that there is a high correlation between the Recall(diff) values
  and both the average pairwise percentage of common courses and the average
  pairwise degree similarity among students of these majors (correlation values
  of 0.47 and 0.5 for \grppopplusminus, and 0.47 and 0.38 for CKRM+SVD(+-),
  respectively). This implies that, as the percentage of common courses and
  degree similarity between pairs of students decrease, accurate course
  recommendation becomes more difficult, since there is more variability in the
  set of courses taken as well as their sequencing. The nursing major, where
  \grppopplusminus\ significantly outperforms CKRM+SVD(+-) has the highest
  average pairwise percentage of common courses, $\sim$76\%, as well as the
  highest average pairwise degree similarity, $\sim$0.86, compared to all other
  majors. This implies that the nursing major is the most restricted major and
  that students tend to follow highly similar degree plans and take very
  similar courses at each academic level. The group popularity ranking in this
  case can easily outperform other recommendation methods.}

\begin{figure}[ht]
  \centering
    \subfloat[Recommendation accuracy per student type.\label{fig:recall-diff-per-student-type}]{
       \includegraphics[width=0.48\textwidth]{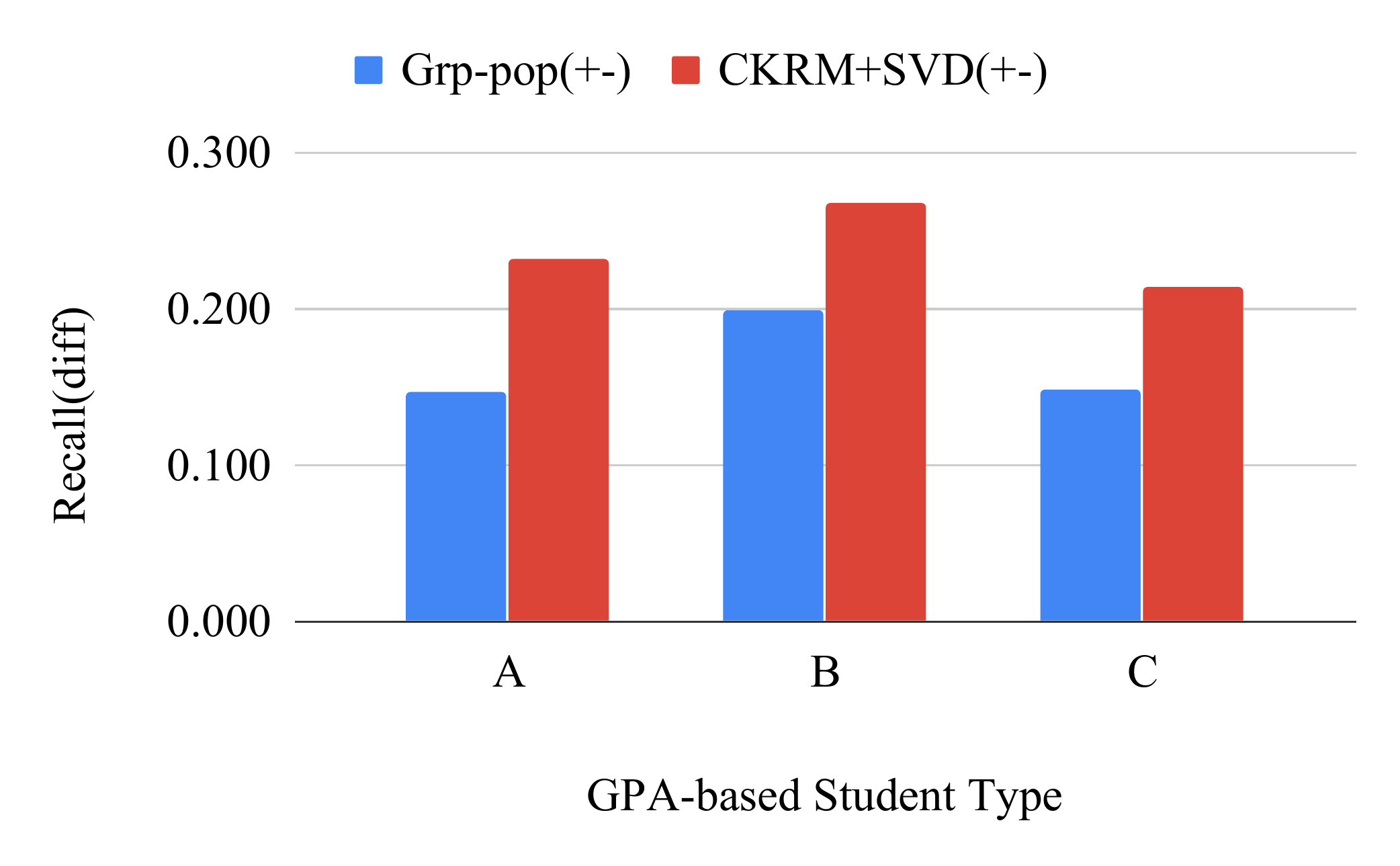}
     }
     \hfill
     \subfloat[Recommendation accuracy per cohort.\label{fig:recall-diff-per-cohort}]{
      \includegraphics[width=0.48\textwidth]{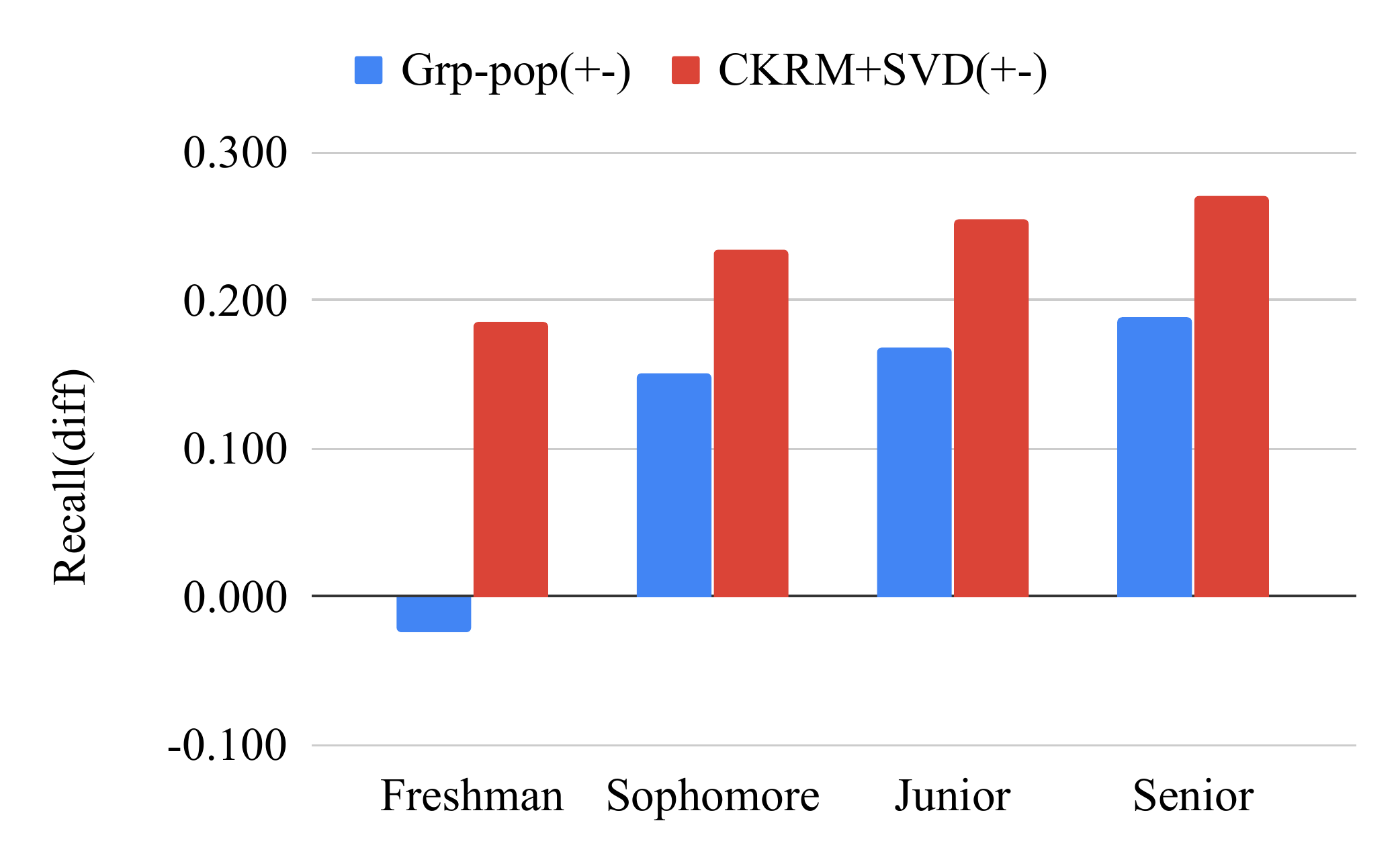} }
    \hfill
    \caption{\label{fig:per-group-analysis}Recommendation accuracy on different
      student sub-groups.}
\end{figure}

\begin{table}[ht]
  \centering
  \caption{Average pairwise degree similarity between different pairs of
    GPA-based student groups.}
  \label{tbl:deg-sim-gpa-based}
  \begin{scriptsize}
    \begin{tabular}{lr}
      \toprule
      \multicolumn{1}{c}{Student Pair} & \multicolumn{1}{c}{Degree Similarity}  \\
      \midrule
      A-B & 0.597 \\
      A-C & 0.535 \\
      B-C & 0.534 \\
      \bottomrule
    \end{tabular}
    \begin{minipage}{0.9\textwidth}
      The column ``Student Pair'' denotes the GPA type of the pair of students
      whose degree similarity was computed.
    \end{minipage}
  \end{scriptsize}
\end{table}

\red{\subsubsection{Analysis on Different Student Groups}}

\red{Figure~\ref{fig:per-group-analysis} shows the recommendation accuracy, in
  terms of Recall(diff), for \grppopplusminus\ and CKRM+SVD(+-) across
  different student sub-groups.}

\red{Figure~\ref{fig:recall-diff-per-student-type} shows the recommendation
  accuracy among different GPA-based student types, A vs B vs C. We notice
  that, first, CKRM+SVD(+-) outperforms \grppopplusminus\ for all student
  groups. Second, we found that CKRM+SVD(+-) achieves the highest Recall(diff)
  for the type-B students, followed by type-A, and then by type-C. This could
  be due to the following reasons. After analyzing the training data, we found
  that the type-A and type-B students constitute $\sim$96\% of the student
  population. After analyzing the average pairwise percentage of common courses
  and degree similarity among each GPA-based groups of students, as well as
  among pairs of different GPA-based groups, we found that type-C students
  follow more diverse sequencing for their degree plans that type-A or type-B
  students, as illustrated in Table~\ref{tbl:deg-sim-gpa-based}, while there
  was no difference among the different groups in the average pairwise
  percentage of common courses. As discussed in
  Section~\ref{sec:res:analysis:majors}, there is a high correlation between
  the pairwise degree similarity and the recommendation accuracy. Since there
  is no enough training data for the type-C students to learn their sequencing
  of the courses, this can explain why the recommendation accuracy for them was
  the lowest.}

\red{Figure~\ref{fig:recall-diff-per-cohort} shows the recommendation accuracy
  among different student sub-groups based on their academic level. At the
  University of Minnesota, there are four academic levels, based on the number
  of both earned and transferred credits by the beginning of the semester: (1)
  freshman ($\le$ 30 credits), (2) sophomore ($> 30$ and $\le 60$ credits), (3)
  junior ($> 60$ and $\le 90$ credits), and senior ($> 90$ credits). First, we
  can notice that CKRM+SVD(+-) significantly outperforms \grppopplusminus\
  across all student groups. Second we see that, as the student's academic
  level increases, and hence he/she has spent more years at the university and
  took more courses, both methods tend to achieve more accurate
  recommendations. This can be due to the following reasons. First, since we
  filter out the courses that have been previously taken by the student before
  making recommendations (see Section~\ref{sec:eval:metrics}), this means that
  as the student's academic level increases, there is a smaller number of
  candidate courses from which the recommendations are to be made. Second, for
  CKRM+SVD(+-), as the student takes more courses, his/her implicit profile
  that is computed by aggregating the embeddings of the previously-taken
  courses becomes more accurate.}

\red{\section{Characteristics of Recommended Courses}}

\red{An important question to any recommendation method is what the
  characteristics of the recommendations are. In this section, we study two
  important characteristics for the recommended courses; (i) the difficulty of
  courses (Section~\ref{sec:res:characteristics:difficulty}), and (ii) their
  popularity (Section~\ref{sec:res:characteristics:popularity}) (RQ7)}.

\red{\subsection{Course Difficulty}}
\label{sec:res:characteristics:difficulty}

\red{As our proposed grade-aware recommendation methods are trained to
  recommend courses that help students maintain or improve their GPA, these
  methods can be prone to recommending more easier courses in which students
  usually achieve high grades. Here, we investigate whether this happens in our
  recommendations or not. Table~\ref{tbl:difficulty-rec-courses} shows the
  grade statistics of all courses, as well as the courses recommended by all
  variations of grade-unaware and grade-aware SVD variations. The mean grade is
  3.5 for all courses, while for the recommended courses, it is 3.24, 3.4, and
  3.56, for SVD(++), SVD(+) and SVD(+-), respectively. These statistics show
  that the grade-aware SVD approaches tend to only slightly favor easier
  courses in their recommendations than the grade-unaware SVD approach.}

\begin{table}[h]
  \centering
  \caption{Statistics for the grades of all and recommended courses.}
  \label{tbl:difficulty-rec-courses}
  \begin{scriptsize}
    \begin{tabular}{lrrr}
      \toprule
      \multicolumn{1}{c}{Course Set} & \multicolumn{1}{c}{Mean} &
                                                                  \multicolumn{1}{c}{Median}
      & \multicolumn{1}{c}{Std. Dev.} \\
      \midrule
      All & 3.50 & 3.61 & 0.51 \\
      SVD(++) & 3.24 & 3.24 & 0.27 \\
      SVD(+) & 3.40 & 3.40 & 0.24 \\
      SVD(+-) & 3.56 & 3.55 & 0.20 \\
      \bottomrule
    \end{tabular}
  \end{scriptsize}
\end{table}

\red{\subsection{Course Popularity}}
\label{sec:res:characteristics:popularity}

\red{Since the university administrators need to make sure that students are
  enrolled in courses with different popularity, as there is a capacity for
  each course and classroom, course popularity is an important factor for
  course recommendations.}

\begin{figure}[ht]
  \centering
  \label{fig:course-popularity}
     \includegraphics[width=0.7\textwidth]{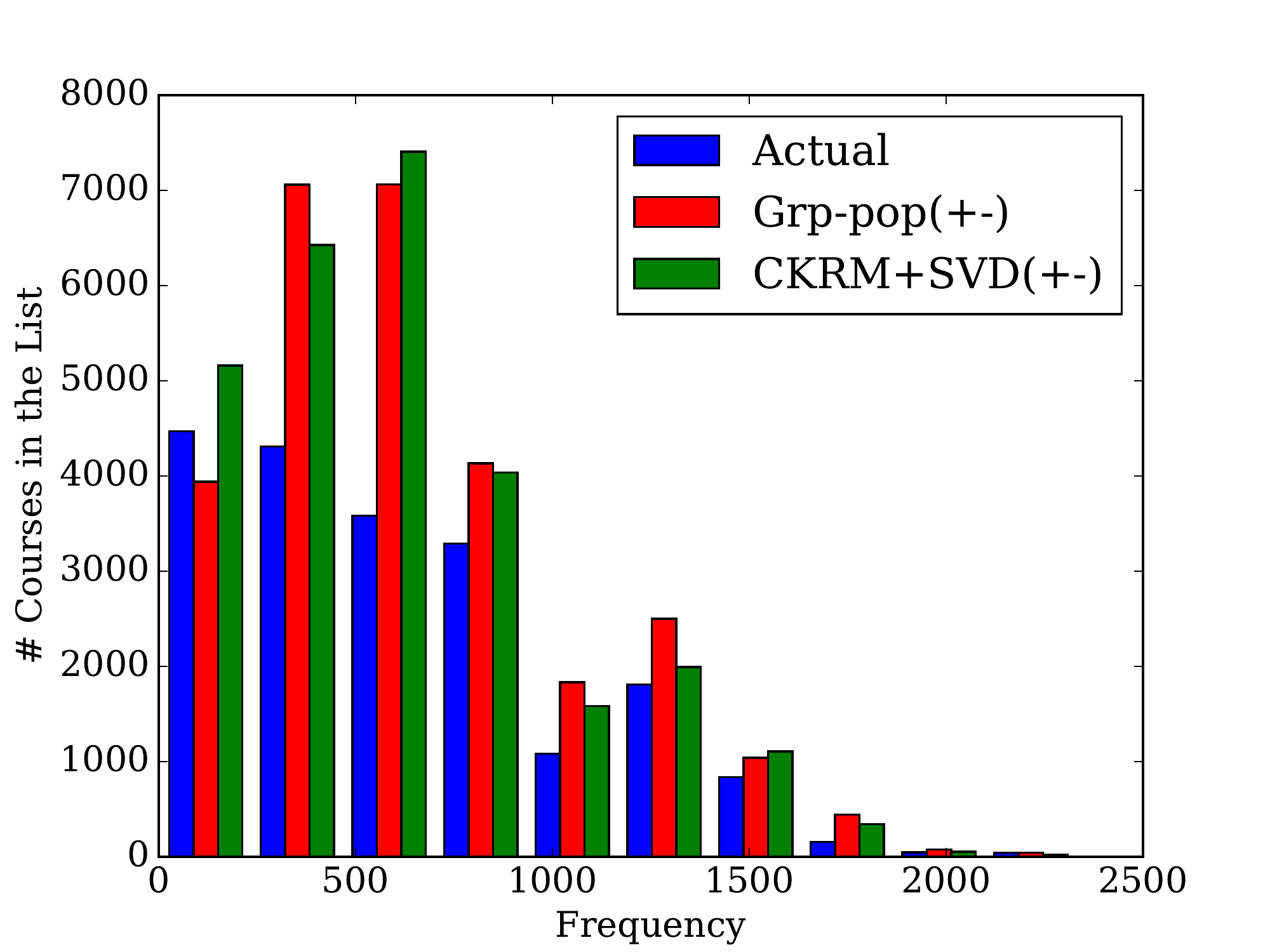}
     \caption{Popularity of the actual good courses, as well as courses
       recommended by \grppopplusminus\ and CKRM+SVD(+-).}
\end{figure}

\red{We also analyze the results of our models in terms of the popularity of
  the courses they recommend. Figure~\ref{fig:course-popularity} shows the
  frequency of the actual good courses in the test set, as well as the
  frequency of the good courses recommended by both \grppopplusminus\ and
  CKRM+SVD(+-)\footnote{Remember that we recommend $n_{(s,t)}$ courses, which
    is the total number of (good and bad) courses taken by student $s$ in term
    $t$ (see Section~\ref{sec:eval:metrics}), so the number of recommendations
    can be higher than the number of actual good courses.}.}

\red{The figure shows that both \grppopplusminus\ and CKRM+SVD(+-) recommend
  courses with different popularity\footnote{Since we use a filtering technique
    before making recommendations, \grppopplusminus\ can recommend courses with
    little popularity (see Section~\ref{sec:eval:metrics})}, similar to the
  actual good courses taken by students. Comparing CKRM+SVD(+-) to
  \grppopplusminus, we can notice that, \grppopplusminus\ tends to recommend a
  higher number of the more popular courses, while CKRM+SVD(+-) recommends more
  of the less popular ones, which can be considered a major benefit for the
  latter method.}

\section{Discussions and Conclusions}
\label{sec:concl}

In this paper, we proposed grade-aware course recommendation approaches for
solving the course recommendation problem. The proposed approach aims to
recommend to students good courses on which the student's expected grades will
maintain or improve their overall GPA. We proposed two different approaches for
solving the grade-aware course recommendation problem. The first approach ranks
the courses by using an objective function that differentiates between
sequences of courses that are expected to increase or decrease a student's
GPA. The second approach combines the grades predicted by grade prediction
methods in order to improve the rankings produced by course recommendation
methods. To obtain course rankings in the first approach, we adapted two
widely-known representation learning techniques; one that uses the linear
Singular Value Decomposition model, while the other uses log-linear neural
network based models.

We conducted an extensive set of experiments on a large dataset obtained from
23 different majors at the University of Minnesota. The results showed that:
(i) the proposed grade-aware course recommendation approaches outperform
grade-unaware recommendation methods in recommending more courses that increase
the students' GPA and fewer courses that decrease it; (ii) the proposed
representation learning based approaches outperform competing approaches for
grade-aware course recommendation; and (iii) the approaches {\blue that utilize
  both the good and bad courses and differentiates between them} achieve
comparable performance to combining grade prediction with the approaches {\blue
  that either utilize the good courses only, or those that differentiate
  between good and bad courses}.

\red{We also provided an in-depth analysis of the recommendation accuracy
  across different majors and student groups. We found that our proposed
  approaches consistently outperformed the best baseline method across these
  majors and groups. We also analyzed the characteristics of the
  recommendations in terms of course difficulty and popularity. We found that
  our proposed grade-aware course recommendation approaches are not prone to
  recommending easy courses, and that they recommend courses with high and low
  popularity in a similar manner. This shows the effectiveness of our proposed
  grade-aware approaches for course recommendation.}

\red{Time-to-degree is another important factor for academic success, which is
  the number of years or terms that the student enrolls in to finish his/her
  degree. An interesting research direction would be to investigate the effect
  of our recommendations on the time-to-degree, and accordingly, develop
  recommendation approaches that considers both the student's GPA and
  time-to-degree.}

\section*{Acknowledgement}

We would like to thank the anonymous reviewers for their valuable feedback on
the original manuscript. This work was supported in part by NSF (1447788,
1704074, 1757916, 1834251), Army Research Office (W911NF1810344), Intel Corp,
and the Digital Technology Center at the University of Minnesota. Access to
research and computing facilities was provided by the Digital Technology Center
and the Minnesota Supercomputing Institute, \url{http://www.msi.umn.edu}.


\bibliographystyle{acmtrans}

\bibliography{refs}

\begin{thebibliography}{}

\bibitem[\protect\citeauthoryear{Backenk{\"o}hler, Scherzinger, Singla, and
  Wolf}{Backenk{\"o}hler et~al\mbox{.}}{2018}]{backenkohler2018data}
{\sc Backenk{\"o}hler, M.}, {\sc Scherzinger, F.}, {\sc Singla, A.}, {\sc and}
  {\sc Wolf, V.} 2018.
\newblock Data-driven approach towards a personalized curriculum.
\newblock In {\em Proceedings of the 11th International Conference on
  Educational Data Mining}. 246--251.

\bibitem[\protect\citeauthoryear{Bell, Koren, and Volinsky}{Bell
  et~al\mbox{.}}{2007}]{bell2007modeling}
{\sc Bell, R.}, {\sc Koren, Y.}, {\sc and} {\sc Volinsky, C.} 2007.
\newblock Modeling relationships at multiple scales to improve accuracy of
  large recommender systems.
\newblock In {\em Proceedings of the 13th ACM SIGKDD International Conference
  on Knowledge Discovery and Data Mining}. KDD '07. ACM, New York, NY, USA,
  95--104.

\bibitem[\protect\citeauthoryear{Bendakir and A{\"\i}meur}{Bendakir and
  A{\"\i}meur}{2006}]{bendakir2006using}
{\sc Bendakir, N.} {\sc and} {\sc A{\"\i}meur, E.} 2006.
\newblock Using association rules for course recommendation.
\newblock In {\em Proceedings of the AAAI Workshop on Educational Data Mining}.
  Vol.~3.

\bibitem[\protect\citeauthoryear{Bhumichitr, Channarukul, Saejiem,
  Jiamthapthaksin, and Nongpong}{Bhumichitr
  et~al\mbox{.}}{2017}]{bhumichitr2017recommender}
{\sc Bhumichitr, K.}, {\sc Channarukul, S.}, {\sc Saejiem, N.}, {\sc
  Jiamthapthaksin, R.}, {\sc and} {\sc Nongpong, K.} 2017.
\newblock Recommender systems for university elective course recommendation.
\newblock In {\em Computer Science and Software Engineering (JCSSE), 2017 14th
  International Joint Conference on}. IEEE, 1--5.

\bibitem[\protect\citeauthoryear{Braxton, Hirschy, and McClendon}{Braxton
  et~al\mbox{.}}{2011}]{braxton2011understanding}
{\sc Braxton, J.~M.}, {\sc Hirschy, A.~S.}, {\sc and} {\sc McClendon, S.~A.}
  2011.
\newblock {\em Understanding and Reducing College Student Departure: ASHE-ERIC
  Higher Education Report, Volume 30, Number 3}. Vol.~16.
\newblock John Wiley \& Sons.

\bibitem[\protect\citeauthoryear{Chen, Moore, Turnbull, and Joachims}{Chen
  et~al\mbox{.}}{2012}]{chen2012playlist}
{\sc Chen, S.}, {\sc Moore, J.~L.}, {\sc Turnbull, D.}, {\sc and} {\sc
  Joachims, T.} 2012.
\newblock Playlist prediction via metric embedding.
\newblock In {\em Proceedings of the 18th ACM SIGKDD international conference
  on Knowledge discovery and data mining}. ACM, 714--722.

\bibitem[\protect\citeauthoryear{Cucuringu, Marshak, Montag, and
  Rombach}{Cucuringu et~al\mbox{.}}{2017}]{cucuringu2017rank}
{\sc Cucuringu, M.}, {\sc Marshak, C.~Z.}, {\sc Montag, D.}, {\sc and} {\sc
  Rombach, P.} 2017.
\newblock Rank aggregation for course sequence discovery.
\newblock In {\em International Workshop on Complex Networks and their
  Applications}. Springer, 139--150.

\bibitem[\protect\citeauthoryear{Elbadrawy and Karypis}{Elbadrawy and
  Karypis}{2016}]{elbadrawy2016domain}
{\sc Elbadrawy, A.} {\sc and} {\sc Karypis, G.} 2016.
\newblock Domain-aware grade prediction and top-n course recommendation.
\newblock In {\em Proceedings of the 10th ACM Conference on Recommender
  Systems}. ACM, 183--190.

\bibitem[\protect\citeauthoryear{Elbadrawy, Studham, and Karypis}{Elbadrawy
  et~al\mbox{.}}{2015}]{elbadrawy2015collaborative}
{\sc Elbadrawy, A.}, {\sc Studham, R.~S.}, {\sc and} {\sc Karypis, G.} 2015.
\newblock Collaborative multi-regression models for predicting students'
  performance in course activities.
\newblock In {\em Proceedings of the 5th International Learning Analytics and
  Knowledge Conference}.

\bibitem[\protect\citeauthoryear{Golub and Reinsch}{Golub and
  Reinsch}{1970}]{golub1970singular}
{\sc Golub, G.~H.} {\sc and} {\sc Reinsch, C.} 1970.
\newblock Singular value decomposition and least squares solutions.
\newblock {\em Numerische mathematik\/}~{\em 14,\/}~5, 403--420.

\bibitem[\protect\citeauthoryear{Gonz{\'a}lez-Brenes and
  Mostow}{Gonz{\'a}lez-Brenes and Mostow}{2012}]{gonzalez2012dynamic}
{\sc Gonz{\'a}lez-Brenes, J.~P.} {\sc and} {\sc Mostow, J.} 2012.
\newblock Dynamic cognitive tracing: Towards unified discovery of student and
  cognitive models.
\newblock {\em EDM\/}.

\bibitem[\protect\citeauthoryear{Grbovic, Radosavljevic, Djuric, Bhamidipati,
  Savla, Bhagwan, and Sharp}{Grbovic et~al\mbox{.}}{2015}]{grbovic2015commerce}
{\sc Grbovic, M.}, {\sc Radosavljevic, V.}, {\sc Djuric, N.}, {\sc Bhamidipati,
  N.}, {\sc Savla, J.}, {\sc Bhagwan, V.}, {\sc and} {\sc Sharp, D.} 2015.
\newblock E-commerce in your inbox: Product recommendations at scale.
\newblock In {\em Proceedings of the 21th ACM SIGKDD International Conference
  on Knowledge Discovery and Data Mining}. ACM, 1809--1818.

\bibitem[\protect\citeauthoryear{Grover and Leskovec}{Grover and
  Leskovec}{2016}]{grover2016node2vec}
{\sc Grover, A.} {\sc and} {\sc Leskovec, J.} 2016.
\newblock node2vec: Scalable feature learning for networks.
\newblock In {\em Proceedings of the 22nd ACM SIGKDD international conference
  on Knowledge discovery and data mining}. ACM, 855--864.

\bibitem[\protect\citeauthoryear{Hagemann, OâMahony, and Smyth}{Hagemann
  et~al\mbox{.}}{2018}]{hagemann2018module}
{\sc Hagemann, N.}, {\sc OâMahony, M.~P.}, {\sc and} {\sc Smyth, B.} 2018.
\newblock Module advisor: Guiding students with recommendations.
\newblock In {\em International Conference on Intelligent Tutoring Systems}.
  Springer, 319--325.

\bibitem[\protect\citeauthoryear{Hershkovitz, Gowda, and Corbett}{Hershkovitz
  et~al\mbox{.}}{2013}]{Hershkovitz2013predicting}
{\sc Hershkovitz, A.}, {\sc Gowda, S.~M.}, {\sc and} {\sc Corbett, A.~T.} 2013.
\newblock Predicting future learning better using quantitative analysis of
  moment-by-moment learning.
\newblock In {\em EDM}.

\bibitem[\protect\citeauthoryear{Hu and Rangwala}{Hu and
  Rangwala}{2018}]{hu2018course}
{\sc Hu, Q.} {\sc and} {\sc Rangwala, H.} 2018.
\newblock Course-specific markovian models for grade prediction.
\newblock In {\em Pacific-Asia Conference on Knowledge Discovery and Data
  Mining}. Springer, 29--41.

\bibitem[\protect\citeauthoryear{Huang, Socher, Manning, and Ng}{Huang
  et~al\mbox{.}}{2012}]{huang2012improving}
{\sc Huang, E.~H.}, {\sc Socher, R.}, {\sc Manning, C.~D.}, {\sc and} {\sc Ng,
  A.~Y.} 2012.
\newblock Improving word representations via global context and multiple word
  prototypes.
\newblock In {\em Proceedings of the 50th Annual Meeting of the Association for
  Computational Linguistics: Long Papers-Volume 1}. Association for
  Computational Linguistics, 873--882.

\bibitem[\protect\citeauthoryear{Hwang and Su}{Hwang and
  Su}{2015}]{hwang2015unified}
{\sc Hwang, C.-S.} {\sc and} {\sc Su, Y.-C.} 2015.
\newblock Unified clustering locality preserving matrix factorization for
  student performance prediction.
\newblock {\em IAENG Int. J. Comput. Sci\/}.

\bibitem[\protect\citeauthoryear{Kena, Hussar, McFarland, de~Brey,
  Musu-Gillette, Wang, Zhang, Rathbun, Wilkinson-Flicker, Diliberti,
  et~al\mbox{.}}{Kena et~al\mbox{.}}{2016}]{kena2016condition}
{\sc Kena, G.}, {\sc Hussar, W.}, {\sc McFarland, J.}, {\sc de~Brey, C.}, {\sc
  Musu-Gillette, L.}, {\sc Wang, X.}, {\sc Zhang, J.}, {\sc Rathbun, A.}, {\sc
  Wilkinson-Flicker, S.}, {\sc Diliberti, M.}, {\sc et~al\mbox{.}} 2016.
\newblock The condition of education 2016. nces 2016-144.
\newblock {\em National Center for Education Statistics\/}.

\bibitem[\protect\citeauthoryear{Koren}{Koren}{2008}]{koren2008factorization}
{\sc Koren, Y.} 2008.
\newblock Factorization meets the neighborhood: a multifaceted collaborative
  filtering model.
\newblock In {\em Proceedings of the 14th ACM SIGKDD international conference
  on Knowledge discovery and data mining}. ACM, 426--434.

\bibitem[\protect\citeauthoryear{Lan, Waters, Studer, and Baraniuk}{Lan
  et~al\mbox{.}}{2014}]{lan2014sparse}
{\sc Lan, A.~S.}, {\sc Waters, A.~E.}, {\sc Studer, C.}, {\sc and} {\sc
  Baraniuk, R.~G.} 2014.
\newblock Sparse factor analysis for learning and content analytics.
\newblock {\em The Journal of Machine Learning Research\/}.

\bibitem[\protect\citeauthoryear{Le and Mikolov}{Le and
  Mikolov}{2014}]{le2014distributed}
{\sc Le, Q.} {\sc and} {\sc Mikolov, T.} 2014.
\newblock Distributed representations of sentences and documents.
\newblock In {\em Proceedings of the 31st International Conference on Machine
  Learning (ICML-14)}. 1188--1196.

\bibitem[\protect\citeauthoryear{Lee and Cho}{Lee and
  Cho}{2011}]{lee2011intelligent}
{\sc Lee, Y.} {\sc and} {\sc Cho, J.} 2011.
\newblock An intelligent course recommendation system.
\newblock {\em SmartCR\/}~{\em 1,\/}~1, 69--84.

\bibitem[\protect\citeauthoryear{Meier, Xu, Atan, and Schaar}{Meier
  et~al\mbox{.}}{2015}]{meier2015personalized}
{\sc Meier, Y.}, {\sc Xu, J.}, {\sc Atan, O.}, {\sc and} {\sc Schaar, M. v.~d.}
  2015.
\newblock Personalized grade prediction: A data mining approach.
\newblock In {\em ICDM}.

\bibitem[\protect\citeauthoryear{Mikolov, Chen, Corrado, and Dean}{Mikolov
  et~al\mbox{.}}{2013}]{mikolov2013efficient}
{\sc Mikolov, T.}, {\sc Chen, K.}, {\sc Corrado, G.}, {\sc and} {\sc Dean, J.}
  2013.
\newblock Efficient estimation of word representations in vector space.
\newblock {\em arXiv preprint arXiv:1301.3781\/}.

\bibitem[\protect\citeauthoryear{Mikolov, Sutskever, Chen, Corrado, and
  Dean}{Mikolov et~al\mbox{.}}{2013}]{mikolov2013distributed}
{\sc Mikolov, T.}, {\sc Sutskever, I.}, {\sc Chen, K.}, {\sc Corrado, G.~S.},
  {\sc and} {\sc Dean, J.} 2013.
\newblock Distributed representations of words and phrases and their
  compositionality.
\newblock In {\em Advances in neural information processing systems}.
  3111--3119.

\bibitem[\protect\citeauthoryear{Morsy and Karypis}{Morsy and
  Karypis}{2017}]{morsy2017cumulative}
{\sc Morsy, S.} {\sc and} {\sc Karypis, G.} 2017.
\newblock Cumulative knowledge-based regression models for next-term grade
  prediction.
\newblock In {\em Proceedings of the 2017 SIAM International Conference on Data
  Mining}. SIAM, 552--560.

\bibitem[\protect\citeauthoryear{Morsy and Karypis}{Morsy and
  Karypis}{2019}]{morsy2019study}
{\sc Morsy, S.} {\sc and} {\sc Karypis, G.} 2019.
\newblock A study on curriculum planning and its relationship with graduation
  gpa and time to degree.
\newblock In {\em Proceedings of the 9th International Conference on Learning
  Analytics \& Knowledge}. ACM, 26--35.

\bibitem[\protect\citeauthoryear{Parameswaran, Venetis, and
  Garcia-Molina}{Parameswaran
  et~al\mbox{.}}{2011}]{parameswaran2011recommendation}
{\sc Parameswaran, A.}, {\sc Venetis, P.}, {\sc and} {\sc Garcia-Molina, H.}
  2011.
\newblock Recommendation systems with complex constraints: A course
  recommendation perspective.
\newblock {\em ACM TOIS\/}~{\em 29,\/}~4, 20.

\bibitem[\protect\citeauthoryear{Parameswaran and Garcia-Molina}{Parameswaran
  and Garcia-Molina}{2009}]{parameswaran2009recommendations}
{\sc Parameswaran, A.~G.} {\sc and} {\sc Garcia-Molina, H.} 2009.
\newblock Recommendations with prerequisites.
\newblock In {\em Proceedings of the third ACM conference on Recommender
  systems}. ACM, 353--356.

\bibitem[\protect\citeauthoryear{Parameswaran, Garcia-Molina, and
  Ullman}{Parameswaran et~al\mbox{.}}{2010}]{parameswaran2010evaluating}
{\sc Parameswaran, A.~G.}, {\sc Garcia-Molina, H.}, {\sc and} {\sc Ullman,
  J.~D.} 2010.
\newblock Evaluating, combining and generalizing recommendations with
  prerequisites.
\newblock In {\em Proceedings of the 19th ACM international conference on
  Information and knowledge management}. ACM, 919--928.

\bibitem[\protect\citeauthoryear{Parameswaran, Koutrika, Bercovitz, and
  Garcia-Molina}{Parameswaran et~al\mbox{.}}{2010}]{parameswaran2010recsplorer}
{\sc Parameswaran, A.~G.}, {\sc Koutrika, G.}, {\sc Bercovitz, B.}, {\sc and}
  {\sc Garcia-Molina, H.} 2010.
\newblock Recsplorer: recommendation algorithms based on precedence mining.
\newblock In {\em Proceedings of the 2010 ACM SIGMOD International Conference
  on Management of data}. ACM, 87--98.

\bibitem[\protect\citeauthoryear{Pardos, Fan, and Jiang}{Pardos
  et~al\mbox{.}}{2019}]{pardos2019connectionist}
{\sc Pardos, Z.~A.}, {\sc Fan, Z.}, {\sc and} {\sc Jiang, W.} 2019.
\newblock Connectionist recommendation in the wild: on the utility and
  scrutability of neural networks for personalized course guidance.
\newblock {\em User Modeling and User-Adapted Interaction\/}, 1--39.

\bibitem[\protect\citeauthoryear{Paterek}{Paterek}{2007}]{paterek2007improving}
{\sc Paterek, A.} 2007.
\newblock Improving regularized singular value decomposition for collaborative
  filtering.
\newblock In {\em Proceedings of KDD cup and workshop}. Vol. 2007. 5--8.

\bibitem[\protect\citeauthoryear{Pennington, Socher, and Manning}{Pennington
  et~al\mbox{.}}{2014}]{pennington2014glove}
{\sc Pennington, J.}, {\sc Socher, R.}, {\sc and} {\sc Manning, C.} 2014.
\newblock Glove: Global vectors for word representation.
\newblock In {\em Proceedings of the 2014 conference on empirical methods in
  natural language processing (EMNLP)}. 1532--1543.

\bibitem[\protect\citeauthoryear{Perozzi, Al-Rfou, and Skiena}{Perozzi
  et~al\mbox{.}}{2014}]{perozzi2014deepwalk}
{\sc Perozzi, B.}, {\sc Al-Rfou, R.}, {\sc and} {\sc Skiena, S.} 2014.
\newblock Deepwalk: Online learning of social representations.
\newblock In {\em Proceedings of the 20th ACM SIGKDD international conference
  on Knowledge discovery and data mining}. ACM, 701--710.

\bibitem[\protect\citeauthoryear{Polyzou and Karypis}{Polyzou and
  Karypis}{2016}]{polyzou2016grade}
{\sc Polyzou, A.} {\sc and} {\sc Karypis, G.} 2016.
\newblock Grade prediction with course and student specific models.
\newblock In {\em PAKDD}. Springer.

\bibitem[\protect\citeauthoryear{Reddy, Labutov, and Joachims}{Reddy
  et~al\mbox{.}}{2016}]{reddy2016latent}
{\sc Reddy, S.}, {\sc Labutov, I.}, {\sc and} {\sc Joachims, T.} 2016.
\newblock Latent skill embedding for personalized lesson sequence
  recommendation.
\newblock {\em arXiv preprint\/}.

\bibitem[\protect\citeauthoryear{Romero, Ventura, Espejo, and
  Herv{\'a}s}{Romero et~al\mbox{.}}{2008}]{romero2008data}
{\sc Romero, C.}, {\sc Ventura, S.}, {\sc Espejo, P.~G.}, {\sc and} {\sc
  Herv{\'a}s, C.} 2008.
\newblock Data mining algorithms to classify students.
\newblock In {\em EDM}.

\bibitem[\protect\citeauthoryear{Sarwar, Karypis, Konstan, and Riedl}{Sarwar
  et~al\mbox{.}}{2000}]{sarwar2000application}
{\sc Sarwar, B.}, {\sc Karypis, G.}, {\sc Konstan, J.}, {\sc and} {\sc Riedl,
  J.} 2000.
\newblock Application of dimensionality reduction in recommender system a case
  study.
\newblock In {\em Proceeding of WebKDD-2000 Workshop}.

\bibitem[\protect\citeauthoryear{Sweeney, Lester, Rangwala, and Johri}{Sweeney
  et~al\mbox{.}}{2016}]{sweeney2016next}
{\sc Sweeney, M.}, {\sc Lester, J.}, {\sc Rangwala, H.}, {\sc and} {\sc Johri,
  A.} 2016.
\newblock Next-term student performance prediction: A recommender systems
  approach.
\newblock {\em Journal of Educational Data Mining\/}~{\em 8,\/}~1, 22--51.

\bibitem[\protect\citeauthoryear{Tang, Qu, Wang, Zhang, Yan, and Mei}{Tang
  et~al\mbox{.}}{2015}]{tang2015line}
{\sc Tang, J.}, {\sc Qu, M.}, {\sc Wang, M.}, {\sc Zhang, M.}, {\sc Yan, J.},
  {\sc and} {\sc Mei, Q.} 2015.
\newblock Line: Large-scale information network embedding.
\newblock In {\em Proceedings of the 24th International Conference on World
  Wide Web}. International World Wide Web Conferences Steering Committee,
  1067--1077.

\bibitem[\protect\citeauthoryear{Thai-Nghe, Drumond, Horv{\'a}th, and
  Schmidt-Thieme}{Thai-Nghe et~al\mbox{.}}{2012}]{thai2012using}
{\sc Thai-Nghe, N.}, {\sc Drumond, L.}, {\sc Horv{\'a}th, T.}, {\sc and} {\sc
  Schmidt-Thieme, L.} 2012.
\newblock Using factorization machines for student modeling.
\newblock In {\em UMAP Workshops}.

\bibitem[\protect\citeauthoryear{Thai-Nghe, Horv{\'a}th, and
  Schmidt-Thieme}{Thai-Nghe et~al\mbox{.}}{2011}]{thai2011factorization}
{\sc Thai-Nghe, N.}, {\sc Horv{\'a}th, T.}, {\sc and} {\sc Schmidt-Thieme, L.}
  2011.
\newblock Factorization models for forecasting student performance.
\newblock In {\em EDM}.

\bibitem[\protect\citeauthoryear{Wang, Guo, Lan, Xu, Wan, and Cheng}{Wang
  et~al\mbox{.}}{2015}]{wang2015learning}
{\sc Wang, P.}, {\sc Guo, J.}, {\sc Lan, Y.}, {\sc Xu, J.}, {\sc Wan, S.}, {\sc
  and} {\sc Cheng, X.} 2015.
\newblock Learning hierarchical representation model for nextbasket
  recommendation.
\newblock In {\em Proceedings of the 38th International ACM SIGIR conference on
  Research and Development in Information Retrieval}. ACM, 403--412.

\bibitem[\protect\citeauthoryear{Xu, Xing, and Van Der~Schaar}{Xu
  et~al\mbox{.}}{2016}]{xu2016personalized}
{\sc Xu, J.}, {\sc Xing, T.}, {\sc and} {\sc Van Der~Schaar, M.} 2016.
\newblock Personalized course sequence recommendations.
\newblock {\em IEEE Transactions on Signal Processing\/}~{\em 64,\/}~20,
  5340--5352.

\end{thebibliography}

\end{document}